\documentclass[10pt]{iopart}
\usepackage{bm}
\usepackage{graphicx}
\eqnobysec

\newcommand{\beq}{\begin{equation}}
\newcommand{\beqa}{\begin{eqnarray}}
\newcommand{\eeq}{\end{equation}}
\newcommand{\eeqa}{\end{eqnarray}}

\newcommand{\abs}[1]{\vert#1\vert}
\renewcommand{\bar}[1]{{\overline{#1}}}
\newcommand{\braket}[2]{\langle#1\vert#2\rangle}
\newcommand{\cm}{{\rm cm}}
\renewcommand{\d}{{\rm d}}
\newcommand{\eps}{\varepsilon}
\newcommand{\estr}{{\boldsymbol\varepsilon}}
\newcommand{\frat}[2]{{\textstyle{\frac{#1}{#2}}}}
\newcommand{\h}{\widehat}
\newcommand{\half}{{\textstyle{\frac{1}{2}}}}
\newcommand{\ii}{{\rm i}}
\newcommand{\ind}[1]{{^{(#1)}}}
\renewcommand{\lg}{{\rm long}}
\renewcommand{\o}{\omega}
\newcommand{\p}{\psi}
\newcommand{\EE}{{\bf E}}
\newcommand{\F}{{(\rm F)}}
\renewcommand{\H}{{\cal H}}
\newcommand{\Int}{\mathop{\rm Int}\nolimits}
\newcommand{\KK}{{\bf K}}
\newcommand{\La}{\Lambda}
\newcommand{\N}{{\cal N}}

\begin{document}

\title[Quantum centipedes]
{Quantum centipedes: collective dynamics of interacting quantum walkers}

\author{P L Krapivsky$^{1,2}$, J M Luck$^2$ and K Mallick$^2$}

\address{$^1$ Department of Physics, Boston University, Boston, MA 02215, USA}

\address{$^2$ Institut de Physique Th\'eorique, Universit\'e Paris-Saclay, CEA and CNRS,
91191 Gif-sur-Yvette, France}

\begin{abstract}
We consider the quantum centipede
made of $N$ fermionic quantum walkers on the one-dimensional lattice
interacting by means of the simplest of all hard-bound constraints:
the distance between two consecutive fermions is either one or two lattice spacings.
This composite quantum walker spreads ballistically, just as the simple quantum walk.
However, because of the interactions between the internal degrees of freedom,
the distribution of its center-of-mass velocity
displays numerous ballistic fronts in the long-time limit,
corresponding to singularities in the empirical velocity distribution.
The spectrum of the centipede and the corresponding group velocities
are analyzed by direct means for the first few values of $N$.
Some analytical results are obtained for arbitrary $N$
by exploiting an exact mapping of the problem onto a free-fermion system.
We thus derive the maximal velocity describing the ballistic spreading
of the two extremal fronts of the centipede wavefunction,
including its non-trivial value in the large-$N$ limit.
\end{abstract}

\eads{\mailto{pkrapivsky@gmail.com},\mailto{jean-marc.luck@cea.fr},\mailto{kirone.mallick@cea.fr}}

%\vspace*{15pt}\address{\today}

\maketitle

\section{Introduction}

Quantum walks~\cite{Aharonov},
the quantum analogues of classical random walks, play
a promi\-nent role in quantum information theory~\cite{Farhi}.
It has been shown in~\cite{Childs}
that any quantum algorithm can be restated in terms of quantum walks.
These universal objects are the source of fascinating problems mixing wave dynamics,
discrete geometry and probability theory.
Quantum dynamics often results in a counter-intuitive phenomenology:
quantities like hitting times or survival probabilities of a walker are genuinely
different in quantum and classical set-ups; in particular,
a quantum search can be far more efficient
than a classical algorithm (see e.g.~\cite{Kempe,Ambainis,Venegas} for reviews).

Single-particle quantum walks have been realized
in the laboratory, using nuclear magnetic resonance~\cite{Ryan},
trapped ions or atoms~\cite{Schmitz,Zahringer,Karski} and photons~\cite{Schreiber}.
The behavior of a single quantum walker
can be explained by a wave description~\cite{Knight} and
reproduced in an experiment with classical waves~\cite{Perets}.

Non-classical effects become essential if one considers multiple quantum walkers.
Quantum walks of correlated photons have been implemented experimentally
by various groups~\cite{Hillery,Peruzzo,Lahini,Sansoni}.
In such systems, quantum interferences and interactions
lead to entanglement and correlations that can not be accounted for by a classical picture,
triggering thus much interest
among theorists~\cite{Omar,Gamble,Stefanak,Chandrashekar,Andrei}.

A single quantum walker displays a ballistic rather than a diffusive motion.
It spreads over a range of space that grows linearly with time.
Surprisingly, the wavefunction displays sharp maxima near the boundaries
of that allowed range, whereas it is negligibly small beyond this range.
These maxima can be interpreted as ballistic fronts~\cite{Farhi,ToroJML}.
In a recent work~\cite{KLM},
we have investigated the dynamics of bosonic and fermionic bound states
of two interacting continuous-time quantum walkers in one dimension.
The emphasis was on the ballistic spreading of the center-of-mass coordinate.
We have demonstrated the existence of multiple internal ballistic fronts,
corresponding to singularities of the velocity distribution,
besides the two usual extremal ones.
This feature is robust and generic,
regardless of the statistics and of the precise form of the
interaction potential between the two particles.

The aim of the present work is to investigate the center-of-mass dynamics,
and especially the ballistic fronts, displayed by a composite object
made of $N$ fermionic quantum walkers on a one-dimensional lattice,
constrained to remain within a fixed distance $\ell$ from their neighbors.
This problem can be viewed as quantum-mechanical version of the diffusive dynamics of
the $N$-legged molecular spiders that were considered in~\cite{Tibor},
hence the name {\it quantum centipede}.
In this work we focus our attention onto the simplest of
all centipedes, corresponding to $\ell=2$.
In this special situation,
some analytical results can be derived
by exploiting an exact mapping of the problem onto a free-fermion system.

The outline of this paper is as follows.
In section~\ref{summary}
we review known results on one-dimensional continuous-time quantum walks,
both for a single walker and for a pair of interacting walkers.
The fermionic quantum centipede studied in this work is defined in section~\ref{centi}.
In section~\ref{mapping} we map the problem onto an integrable $XX$ Heisenberg spin chain,
which can be reduced to a free-fermion system
and diagonalized by means of a Jordan-Wigner transformation.
Explicit results on the spectrum of the quantum centipede
are presented in section~\ref{explicit}
for the first few values of the fermion number ($N=2$ to $5$).
In section~\ref{vmax} we obtain the maximal spreading velocities $V\ind{N}$
for arbitrary $N$, as well as their limit $V\ind{\infty}$.
Section~\ref{discussion} contains a discussion of our findings.
A derivation of the characteristic equations~(\ref{chnp}),~(\ref{chni})
is given in~\ref{app}.

\section{A summary of earlier results on one and two quantum walkers}
\label{summary}

We consider continuous-time quantum walks on the discrete one-dimensional lattice.
There is no need for an internal degree of freedom (quantum coin),
as would be required for discrete-time dynamics.
We recall some elementary results
for the single quantum walk~\cite{Farhi,Kempe,Ambainis,Venegas,ToroJML},
and then discuss the case of two co-walking particles,
with an emphasis on the ballistic fronts (see~\cite{KLM} and the references therein).

\subsection{The simple quantum walker}

The simple continuous-time quantum walk is modeled by a tight-binding Hamiltonian,
in which the walker hops from a site to a neighboring site.
We denote by $\psi_n(t) = \langle n | \psi(t) \rangle$ the
wavefunction of the particle at site $n$ at time $t$, and use dimensionless units.
The dynamics of the walker is given by
\beq
\ii\,\frac{\d \psi_n(t)}{\d t} = \psi_{n+1}(t) + \psi_{n-1}(t).
\label{eq:1QW}
\eeq

Suppose that the particle is launched from the origin at time $t=0$:
$\psi_n(0) = \delta_{n0}$.
The wavefunction at time $t$ is then given by a Bessel function:
\beq
\psi_n(t) = \ii^{-n} J_n(2t).
\label{Sol:1QW}
\eeq
Asymptotic properties of Bessel functions allow us to analyze the spreading
of the quantum walk in the long-time limit~\cite{ToroJML}.
The asymptotic probability distribution of
the effective velocity $v = n/t$ has a compact support and converges to an `arc-sine~law':
\beq
f(v) = \frac{1}{ \pi \sqrt{4 -v^2}}\qquad(\abs{v}<2).
\eeq
We emphasize that, in contrast
to the classical case, the convergence is in the weak sense:
the probability distribution $|\psi_n(t)|^2$ displays
high-frequency oscillations which must be averaged out to derive
the function $f(v)$~\cite{Baraviera,Grimmett,Gottlieb,Konno,Strauch}.

At late times, the quantum particle is therefore almost surely located in the allowed
region ($\abs{n}<2t$).
A more precise analysis shows that the probabilities $|\psi_n(t)|^2$
display sharp ballistic fronts near the endpoints of the allowed region ($n = \pm 2t$),
with a height scaling as $t^{-2/3}$ and a width scaling as $t^{1/3}$.
The above generic behavior remains unchanged as long as
the initial state is localized in a finite region:
the quantum walker spreads ballistically in the allowed region limited
by ballistic fronts near $n = \pm 2t$, with a forbidden region beyond them.
It is however possible to engineer exceptional initial states,
for which either one or even both fronts are eliminated
by quantum interferences~\cite{KLM}, but these features are non-generic.

The picture changes qualitatively if the particle is allowed to hop
to the next-nearest neighboring sites with a transition amplitude $g$:
\beq
\ii\,\frac{\d \psi_n(t)}{\d t} = \psi_{n+1}(t) + \psi_{n-1}(t)
+ g \left( \psi_{n+2}(t) + \psi_{n-2}(t) \right).
\label{eq:1QWNN}
\eeq
Allowing hopping to second and further neighbors is known to have
far reaching consequences in a variety of situations~\cite{r1,r2,x1,x2}.
For instance, in the case of graphene~\cite{g1,g2},
hopping to second neighbors breaks the chiral symmetry between both sublattices.
In the present case, when $g>1/4$,
the probability distribution of the velocity $v = n/t$ becomes singular at four values.
The quantum walker thus exhibits four fronts:
two external fronts (as above) at $v=\pm V_+$,
and also two internal fronts at $v=\pm V_-$~\cite{KLM}.
If longer range hopping is allowed and if the corresponding hopping amplitudes exceed
critical values, more internal fronts might appear for generic initial conditions.

\subsection{Two co-walking quantum particles}
\label{twow}

The continuous-time quantum walk problem
can be generalized by considering several interacting quantum walkers.
In a recent work~\cite{KLM}, we have investigated the quantum walk performed by
two identical particles interacting either through hard-bound constraints
or by a smooth confining potential.
The statistics of the particles (bosonic or fermionic) turned out to play
an important role in the analysis of the ballistic spreading
of the bound states thus obtained.

We briefly summarize the results of~\cite{KLM}
for the quantum walk of two one-dimensional fermions
interacting by the hard-bound constraint that their distance is at most $\ell$
lattice spacings.
We denote by $n_1=n+m$ and $n_2=n$ the positions of the particles,
so that $n_\cm=n+m/2$ is the center-of-mass coordinate,
whereas $m=n_1-n_2$ is the relative coordinate.
The hard-bound constraint imposes that $|m| \le \ell $.
This fermionic system is described by the wavefunction
\beq
\p_{n,m}(t)=\braket{(n_1,n_2)}{\p(t)}=\braket{(n+m,n)}{\p(t)},
\eeq
which is odd with respect to $m$.
Because fermionic particles can not cross one another in one dimension,
$m$ can be restricted to the range $m=1,\dots,\ell$.
The dynamics is then given by
\beqa
\ii\,\frac{\d\p_{n,m}(t)}{\d t} &=& \p_{n,m-1}(t)+\p_{n+1,m-1}(t)
\nonumber\\
&+&\p_{n-1,m+1}(t)+\p_{n,m+1}(t),
\label{eq:2-body}
\eeqa
with Dirichlet boundary conditions: $\p_{n,0}(t)=\p_{n,\ell+1}(t)=0$.

The exact solution of this two-body problem displays the following features.
The wavefunction again spreads ballistically in the center-of-mass coordinate.
For late times, the components $\p_{n,m}(t)$ of the wavefunction have appreciable
values for a range of $n$ that grows ballistically and symmetrically
with respect to the origin.
The probability distribution $|\p_{n,m}(t)|^2$ of the bound state
in its center-of-mass coordinate
generically exhibits sharp ballistic fronts for $n \approx V_kt$,
where the front velocities read
\beq
V_k = 2\cos\frac{k\pi}{\ell+1}\qquad(k=1,\dots,\ell).
\eeq
The spreading dynamics is therefore characterized by two extremal fronts,
and $\ell -2 $ internal ones for $\ell\ge3$.
The range of the allowed zone is $|n| < Vt$, the maximal spreading velocity being
\beq
V=2\cos\frac{\pi}{\ell+1}.
\eeq
In the limit where the extent of the bound state diverges ($\ell \to \infty$),
the above result approaches the free value $V = 2$, with a $1/\ell^2$ correction.

This picture remains qualitatively unchanged
if the hard-bound constraint is replaced by a smooth confining potential.
The bosonic and fermionic spectra are infinite sequences of dispersive
energy levels, each of which giving rise to a ballistic front.
These spectra have been studied in detail
in the case where the confining potential is homogeneous,
i.e., of the form $W_m = g |m|^\alpha$.
In particular, the maximal spreading velocity
of two-fermion bound states departs from its free value $V=2$ according to
\beq
V\approx2-C^\F\,g^{2/(\alpha+3)}
\eeq
at weak coupling ($g\ll1$),
where the constant $C^\F$ has been determined~\cite{KLM}.

\section{The fermionic quantum centipede}
\label{centi}

We now introduce the system we study in this work.
It is the quantum centipede made of $N$ interacting fermionic quantum walkers
on the one-dimensional lattice.
The interaction is modeled by the simplest of all hard-bound constraints:
the distance between two consecutive fermions is either one or two lattice spacings.
Besides the number $N$ of fermions, the model is entirely parameter-free.

One-dimensional fermions cannot cross each other,
and so the discrete positions of the particles along the chain
can be assumed to be ordered as $x_1<x_2<\dots<x_N$.
We label the state of the quantum centipede by the following variables:

\begin{itemize}

\item
$n=x_1$ denotes the position of the leftmost fermion along the chain,

\item
the internal state of the centipede is described by a string
$\estr=(\eps_1,\dots,\eps_{N-1})$ of $N-1$ binary variables,
with $\eps_j = x_{j+1} - x_j -1=0$ or 1 for $j=1,\dots,N-1$.
We have thus $\eps_j=0$ if fermions $j$ and $j+1$ are adjacent,
while $\eps_j=1$ if they are separated by a single empty site.

\end{itemize}

\begin{figure}[!ht]
\begin{center}
\includegraphics[angle=-90,width=.5\linewidth]{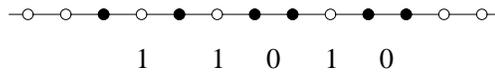}
\caption{\small
A configuration of 6 fermions obeying the hard-bound constraint
and the corresponding string $\estr$ of 5 binary variables.}
\label{config}
\end{center}
\end{figure}

Figure~\ref{config} shows a configuration of 6 fermions
and the corresponding string $\estr$.

The center-of-mass coordinate of the centipede reads
\beq
x_\cm=n+\frac{N-1}{2}+\frac{1}{N}\sum_{j=1}^{N-1}(N-j)\eps_j.
\eeq
We have $x_\cm=n+(N-1)/2$ for the most compact internal state ($\eps_j=0$ for all $j$),
whereas $x_\cm=n+N-1$ for the most extended one ($\eps_j=1$ for all $j$).

The $2^{N-1}$ amplitudes of $N$-body wavefunction
\beq
\psi_n^\estr(t)=\langle x_1,x_2,\dots,x_N | \psi(t) \rangle
\label{psiestr}
\eeq
satisfy coupled continuous-time dynamical equations,
which are analogous to~(\ref{eq:2-body}).
We shall not need to write down these equations explicitly,
except for $N=2$ (see~(\ref{am2})) and $N=3$ (see~(\ref{am3})).
The system is spatially homogeneous, i.e., invariant under discrete translations.
It is therefore convenient to perform a Fourier transform with respect to~$n$ and define
\beq
\h\psi^\estr(q,t) = \sum_n \e^{-\ii qn} \psi_n^\estr(t),
\label{pfou}
\eeq
where the center-of-mass momentum $q$
can be restricted to the first Brillouin zone ($\abs{q}\le\pi$).

The dynamics of the amplitudes~(\ref{pfou}) is governed by an effective dispersive
(i.e., $q$-dependent) Hamiltonian $\H$,
represented by a Hermitian matrix of size $2^{N-1} \times 2^{N-1}$
(see e.g.~(\ref{ham2}),~(\ref{ham3}),~(\ref{ham4})).
The energy spectrum of the centipede therefore has $2^{N-1}$ branches,
i.e., the eigenvalues $\o_a(q)$ of $\H$, with $a=1,\dots,2^{N-1}$.
The corresponding branches of the group velocity read
\beq
v_a(q)=\frac{\d\omega_a(q)}{\d q}.
\label{vdef}
\eeq
We shall be mostly interested in the maximal velocity
\beq
V\ind{N}=\max_{a,q}\abs{v_a(q)},
\label{vmaxdef}
\eeq
describing the ballistic spreading of the two extremal fronts of the wavefunction
in the center-of-mass coordinate, as well as in the internal ballistic fronts,
characterized by all the other stationary values of the group velocity, such that
\beq
\frac{\d v_a(q)}{\d q}=\frac{\d^2\omega_a(q)}{\d q^2}=0.
\label{dvdq}
\eeq

In order to proceed,
we start by noticing that the action of the quantum Hamiltonian $\H$
can be described in purely classical terms.
The string $\estr$ is interpreted as
a classical configuration of particles and holes on an open finite lattice of size $N-1$.
If $\eps_j =1$, site $j$ is occupied by a particle; if $\eps_j =0$, site~$j$ is empty.
The quantum dynamics generated by $\H$ corresponds to the following evolution rules:
\beqa
\hbox{Bulk:}&10 \rightleftharpoons 01\quad &\hbox{with rate 1}.
\nonumber\\
\hbox{Site 1:}&1 \rightarrow 0 &\hbox{with rate } \e^{-\ii q},
\nonumber\\
&0 \rightarrow 1 \quad &\hbox{with rate } \e^{\ii q}.
\nonumber\\
\hbox{Site $N-1$:}\quad &1 \rightarrow 0 &\hbox{with rate 1},
\nonumber\\
&0 \rightarrow 1 &\hbox{with rate 1}.
\label{rules}
\eeqa
These rules are reminiscent
of the symmetric simple exclusion process (SEP) with open boundaries
(see e.g.~\cite{krbbook}).
An equivalence with the SEP had already been put forward
in the classical situation of the molecular spiders and centipedes
investigated in~\cite{Tibor}.
There are however several notable differences
between quantum-mechanical systems such as the present one
and classical stochastic systems such as the SEP:
(i)~The~quantum Hamiltonian $\H$ acts on amplitudes, and not on probabilities.
(ii)~The~transition amplitudes or `rates' are not necessarily positive real numbers.
(iii)~The~system is not equivalent to a classical stochastic process, even for~$q=0$.
The Hamiltonian~$\H$ and the Markov operator for the SEP
have the same non-diagonal elements,
but the Markov operator contains diagonal loss terms,
in order to ensure probability conservation,
whereas the Hamiltonian $\H$ does not have diagonal entries.

\section{Mapping onto a free-fermion system}
\label{mapping}

The energy spectrum of the quantum centipede can be determined, at least formally,
by means of an exact mapping onto an integrable spin chain
and finally onto a free-fermion system.

The Hamiltonian $\H$ that implements the quantum dynamics
(\ref{rules}) can be written, using Pauli matrices, as
\beq
\H = \e^{-\ii q} S_1^{+} + \e^{\ii q} S_1^{-}
+ \sum_{j=1}^{N-2} \! \left( S_{j}^{-} S_{j+1}^{+} + S_{j}^{+} S_{j+1}^{-} \right)
+ S_{N-1}^{+} + S_{N-1}^{-}.
\label{Heffectif}
\eeq
We thus obtain the Hamiltonian of an $XX$ spin chain
with non-diagonal boundary terms~\cite{Nepomechie,Samaj}.
By convention,
$\eps_j=0$ (site $j$ is empty) corresponds to $\uparrow_j$ (spin $j$ is~up),
whereas $\eps_j=1$ (site $j$ is occupied) corresponds to $\downarrow_j$ (spin $j$ is down).
In the local basis
$\{0_j,1_j\} \equiv \{| \uparrow_{j} \rangle, |\downarrow_{j} \rangle\}$,
the Pauli matrices are given by
\beq
S_j^x=\pmatrix{0&1\cr1&0},\quad
S_j^y=\pmatrix{0&-\ii\cr\ii&0},\quad
S_j^z=\pmatrix{1&0\cr0&-1}
\label{Pauli}
\eeq
and the raising and lowering operators $S_j^\pm$ are defined as
\beq
S_j^+=\half(S_j^x+\ii S_j^y)=\pmatrix{0&1\cr0&0},
\ S_j^-=\half(S_j^x-\ii S_j^y)=\pmatrix{0&0\cr1&0}.
\eeq

The Hamiltonian $\H$
can be diagonalized by means of a Jordan-Wigner transfor\-mation~\cite{Lieb}
mapping it onto a free-fermion system.
Because of the boundary terms,~$\H$ is not fully bilinear.
This can be rectified~\cite{Hinrichsen,Birgit}
by adding two auxiliary sites, one at each end of the chain, labeled 0 and $N$.
We thus define a new Hamiltonian $\H_\lg$ on a chain of $N+1$ sites as
\beqa
\H_\lg &=& \e^{-\ii q} S_0^{x} S_1^{+} + \e^{\ii q} S_0^{x} S_1^{-}
+ \sum_{j=1}^{N-2} \left( S_{j}^{-} S_{j+1}^{+} + S_{j}^{+} S_{j+1}^{-} \right)
\nonumber\\
&+& S_{N-1}^{+} S_N^{x} + S_{N-1}^{-} S_N^{x}.
\label{Hlong}
\eeqa
The boundary operators $S_0^{x}$ and $S_N^{x}$ commute with $\H_\lg$.
Hence the eigenstates of~$\H_\lg$ belong to four distinct sectors,
corresponding to the eigenvalues $(\pm 1,\pm 1)$
of the operators $S_0^{x}$ and $S_N^{x}$.
The restriction of $\H_\lg$ to the sector
$(+1,+1)$ coincides with the Hamiltonian $\H$ of~(\ref{Heffectif}).

The Hamiltonian $\H_\lg$ can be diagonalized using
a fermionization procedure, as explained in~\cite{Birgit}.
The operators defined as
\beq
\tau_{j}^{x,y} = \left( \prod_{i=0}^{j-1} S_{i}^z \right) S_{i}^{x,y}
\qquad(j=0,\dots,N)
\label{JW}
\eeq
satisfy the relations
\beq
\{\tau_{j}^\mu, \tau_{k}^\nu \} = 2\delta_{jk}\delta^{\mu\nu}
\qquad(\mu,\nu=x,y).
\label{Clifford}
\eeq
The above anti-commutation relations define a Clifford algebra.
If we rewrite $\H_\lg$ in terms of these operators, we obtain
\beqa
- \H_\lg &=& \ii \cos{q}\; \tau_{0}^y \tau_{1}^x + \ii \sin{q}\; \tau_{0}^y \tau_{1}^y
\nonumber\\
&+&\frac{\ii}{2} \sum_{j=1}^{N-2} \left( \tau_{j}^y \tau_{j+1}^x - \tau_{j}^x \tau_{j+1}^y \right)
+ \ii \tau_{N-1}^y \tau_N^x.
\label{JWHlong}
\eeqa

The last step consists in expressing $\H_\lg$ as a free-fermion Hamiltonian:
\beq
\H_\lg = \sum_{k=0}^N \La_k (2 a_k^\dagger a_k -1).
\label{Hab}
\eeq
To do so, we must find a set of annihilation and creation operators
$a_k$ and $a_k^\dagger$ of fermionic quasiparticles,
satisfying the canonical anti-commutation relations
\beq
\{a_k, a_l^\dagger \} = \delta_{kl},\quad
\{a_k, a_l \} = \{a_k^\dagger, a_l^\dagger \}=0\qquad(k=0,\dots,N).
\label{fermionsab}
\eeq
These quasiparticles are not to be confused
with the original fermionic quantum walkers which constitute the centipede.
The number operators $\N_k = a_k^\dagger a_k$ have eigenvalues 0 and 1.
It follows from~(\ref{Hab}) that the eigenvalues of $\H_\lg$ are given~by
\beq
\o_\lg=\sum_{k=0}^N(\pm\La_k),
\label{ohl}
\eeq
where the sign $\pm$ in front of $\La_k$
depends on whether the $k$-th quasiparticle is present ($\N_k=1$)
or absent ($\N_k=0$).

The quasiparticle operators $a_k$ and $a_k^\dagger$ are obtained from
the Jordan-Wigner operators~$\tau_{j}^{x,y}$ by a Bogoliubov transformation of the form
\beqa
a_k &=& \frac{1}{2} \sum_{j=0}^N
\left( x_{k;j} \, \tau_{j}^x + y_{k;j}\, \tau_{j}^y \right),
\nonumber\\
a_k^\dagger &=& \frac{1}{2} \sum_{j=0}^N
\left( \bar{x}_{k;j} \, \tau_{j}^x + \bar{y}_{k;j} \, \tau_{j}^y \right),
\label{aadag}
\eeqa
where the bar denotes complex conjugation.
The complex coefficients $ (x_{k;j}, y_{k;j})$ are found by
requiring that $a_k$ and $a_k^\dagger$ satisfy the canonical fermionic
anti-commutation relations~(\ref{fermionsab})
and that $\H_\lg$ takes the diagonal form~(\ref{Hab}).
These constraints are implemented by writing
the commutation relations between $\H_\lg$ and $a_k$, $a_k^\dagger$:
\beq
\lbrack \H_\lg, a_k \rbrack = - 2 \La_k a_k,\quad
\lbrack \H_\lg, a_k^\dagger \rbrack = 2 \La_k a_k^\dagger.
\label{eq:Heisenberg}
\eeq
Details are given in~\ref{app}.
The quasiparticle eigenvalues are given by
\beq
\La_k = \cos p_k,
\label{lp}
\eeq
where the discrete values $p_k$ of the internal momentum $p$
satisfy the characteristic equation~(\ref{eq:caract}),
which can be further simplified by dealing separately with even and odd values of~$N$.
We obtain after some algebra
\beqa
N\hbox{ even:}\quad & \sin((N+1)p)-3\sin((N-1)p)=\pm4\sin p\sin q,
\label{chnp}
\\
N\hbox{ odd:}\hfill & \sin((N+1)p)-3\sin((N-1)p)=\pm4\sin p\cos q.
\label{chni}
\eeqa

\section{Explicit results for the first few values of $N$}
\label{explicit}

In this section we present explicit results for the first few values
of the fermion number ($N=2$ to $5$).

\subsection*{$\bullet$ $N=2$}

This is a special case of the more general two-body problem considered in~\cite{KLM},
where the maximal distance between the two quantum walkers is an arbitrary integer~$\ell$.

\begin{figure}[!ht]
\begin{center}
\includegraphics[angle=-90,width=.45\linewidth]{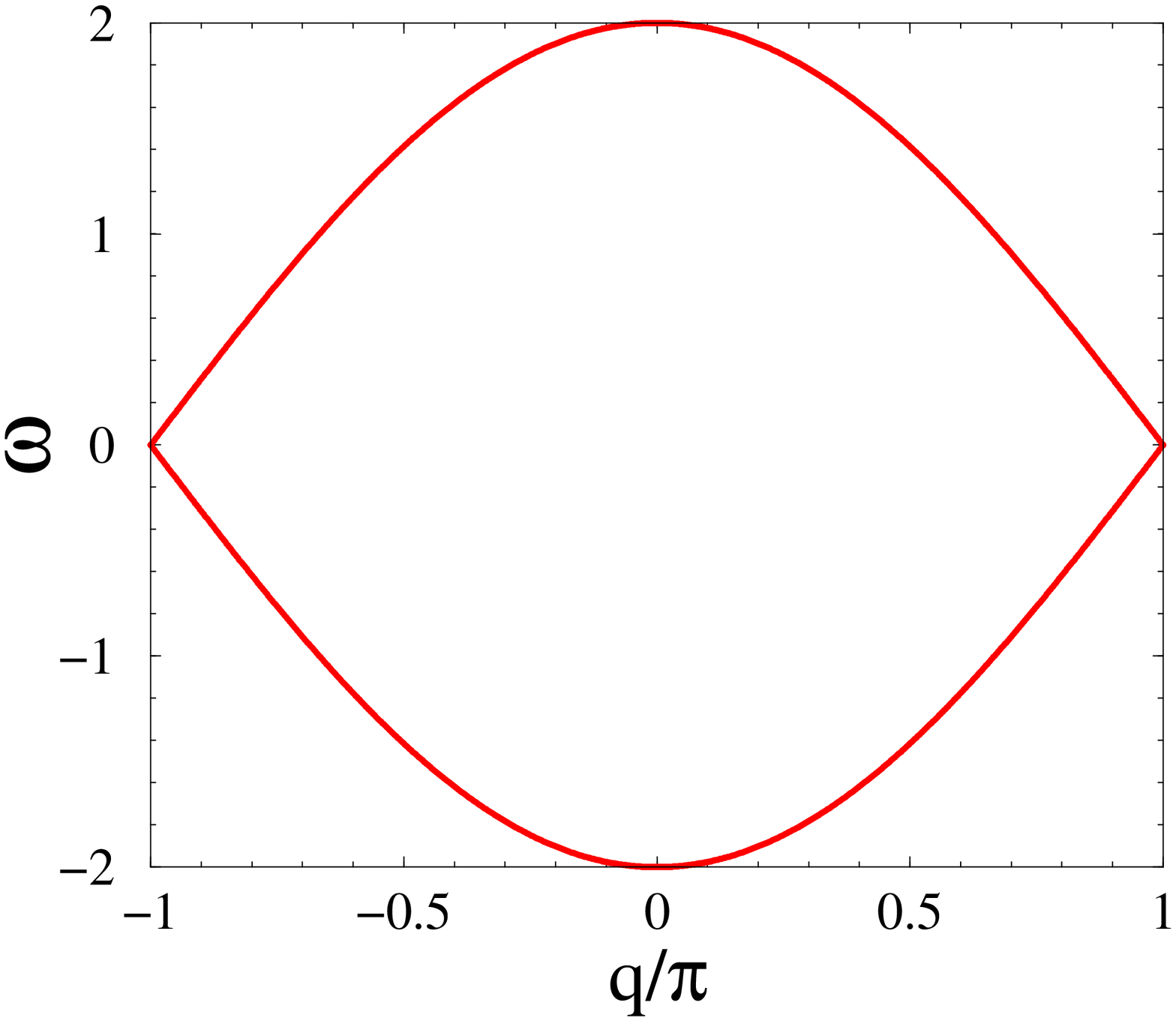}
\includegraphics[angle=-90,width=.45\linewidth]{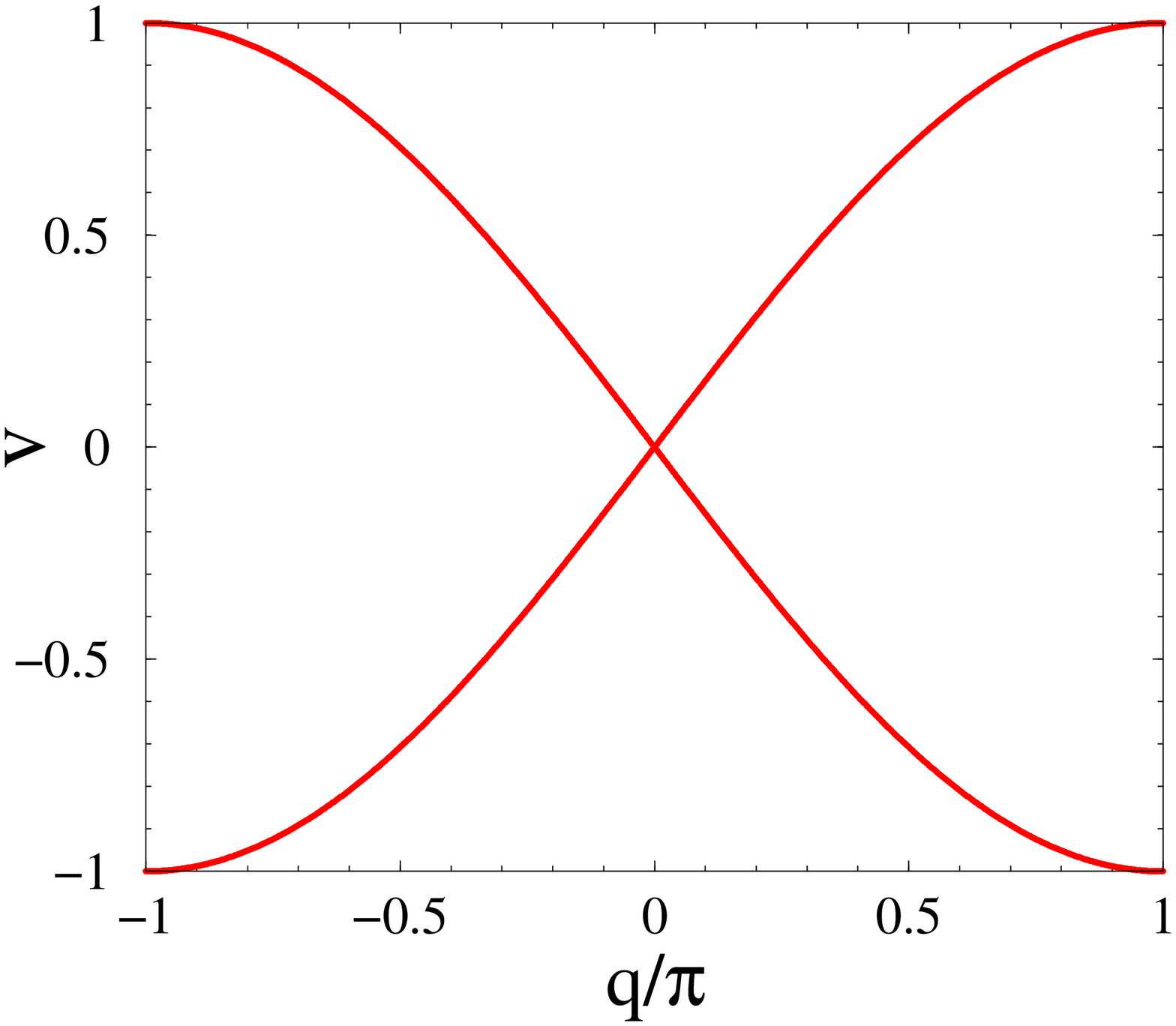}
\caption{\small
Left: energy spectrum of the $N=2$ centipede against $q/\pi$.
Right: associated group velocities.}
\label{wv2}
\end{center}
\end{figure}

With the notation~(\ref{psiestr}),
the amplitudes $\p_n^0$ and $\p_n^1$ obey the equations
\beqa
\ii\,\frac{\d\p_n^0(t)}{\d t} &=& \p_{n-1}^1(t)+\p_n^1(t),
\nonumber\\
\ii\,\frac{\d\p_n^1(t)}{\d t} &=& \p_n^0(t)+\p_{n+1}^0(t),
\label{am2}
\eeqa
which can be viewed as a special case of~(\ref{eq:2-body}).
The corresponding Hamiltonian reads
\beq
\H=\pmatrix{0&1+\e^{\ii q}\cr 1+\e^{-\ii q}&0}.
\label{ham2}
\eeq
We thus readily obtain
\beq
\o_{1,2}=\pm 2\cos\frac{q}{2}.
\label{o2}
\eeq
The associated group velocities read
\beq
v_{1,2}=\mp\sin\frac{q}{2}.
\eeq
In particular, the maximal velocities $\pm V\ind{2}$, with
\beq
V\ind{2}=1,
\label{v2}
\eeq
are reached for $q=\pm\pi$.
There is no other stationary value of the group velocity,
and consequently no internal front besides the extremal ones,
in agreement with the findings of~\cite{KLM} for $\ell=2$, recalled in section~\ref{twow}.
Figure~\ref{wv2} shows plots of the energy spectrum (left)
and of the group velocities (right) against $q/\pi$ in the first Brillouin~zone.

\begin{figure}[!ht]
\begin{center}
\includegraphics[angle=-90,width=.45\linewidth]{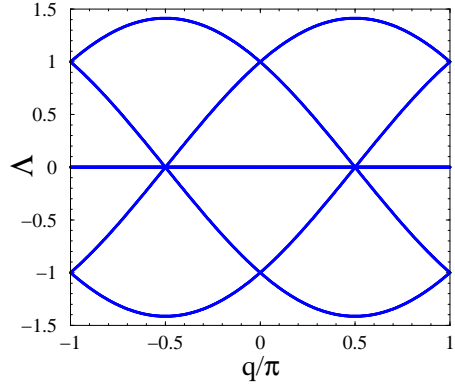}
\caption{\small
Quasiparticle spectrum of the $N=2$ centipede against~$q/\pi$.}
\label{h2}
\end{center}
\end{figure}

Equation~(\ref{chnp}) yields the quasiparticle eigenvalues
\beq
\La_{1,2}=\sqrt{1\pm\sin q}.
\eeq
Figure~\ref{h2} shows the quasiparticle spectrum against $q/\pi$.
The correspondence~(\ref{ohl}) relies on the following identities:
\beq
\o_1=-\o_2=\left\{\matrix{
\La_2-\La_1\quad & (-\pi\le q\le -\pi/2),\cr
\La_1+\La_2\hfill & (-\pi/2\le q\le \pi/2),\cr
\La_1-\La_2\hfill & (\pi/2\le q\le \pi).\hfill}\right.
\eeq

\subsection*{$\bullet$ $N=3$}

\begin{figure}[!ht]
\begin{center}
\includegraphics[angle=-90,width=.45\linewidth]{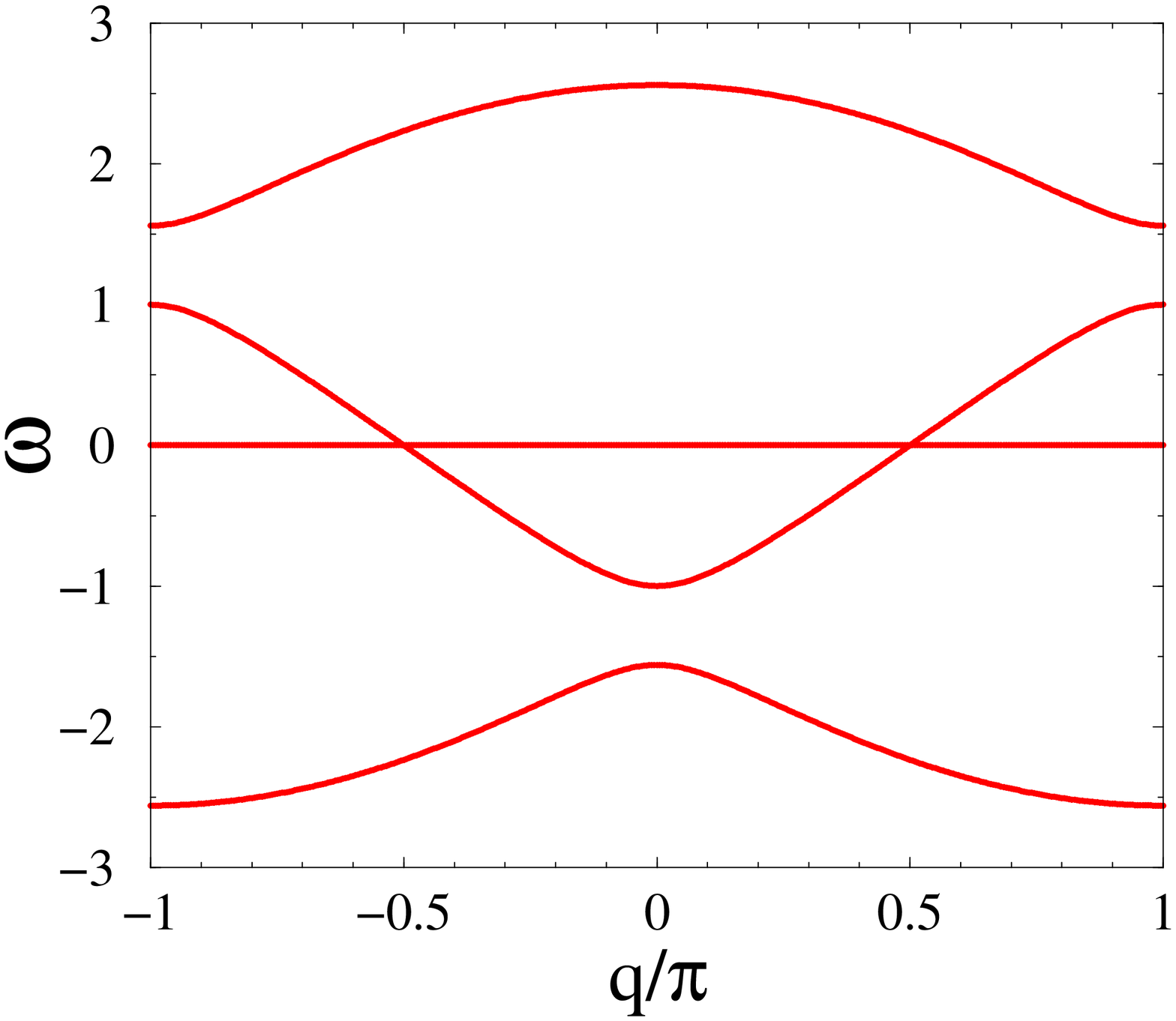}
\includegraphics[angle=-90,width=.45\linewidth]{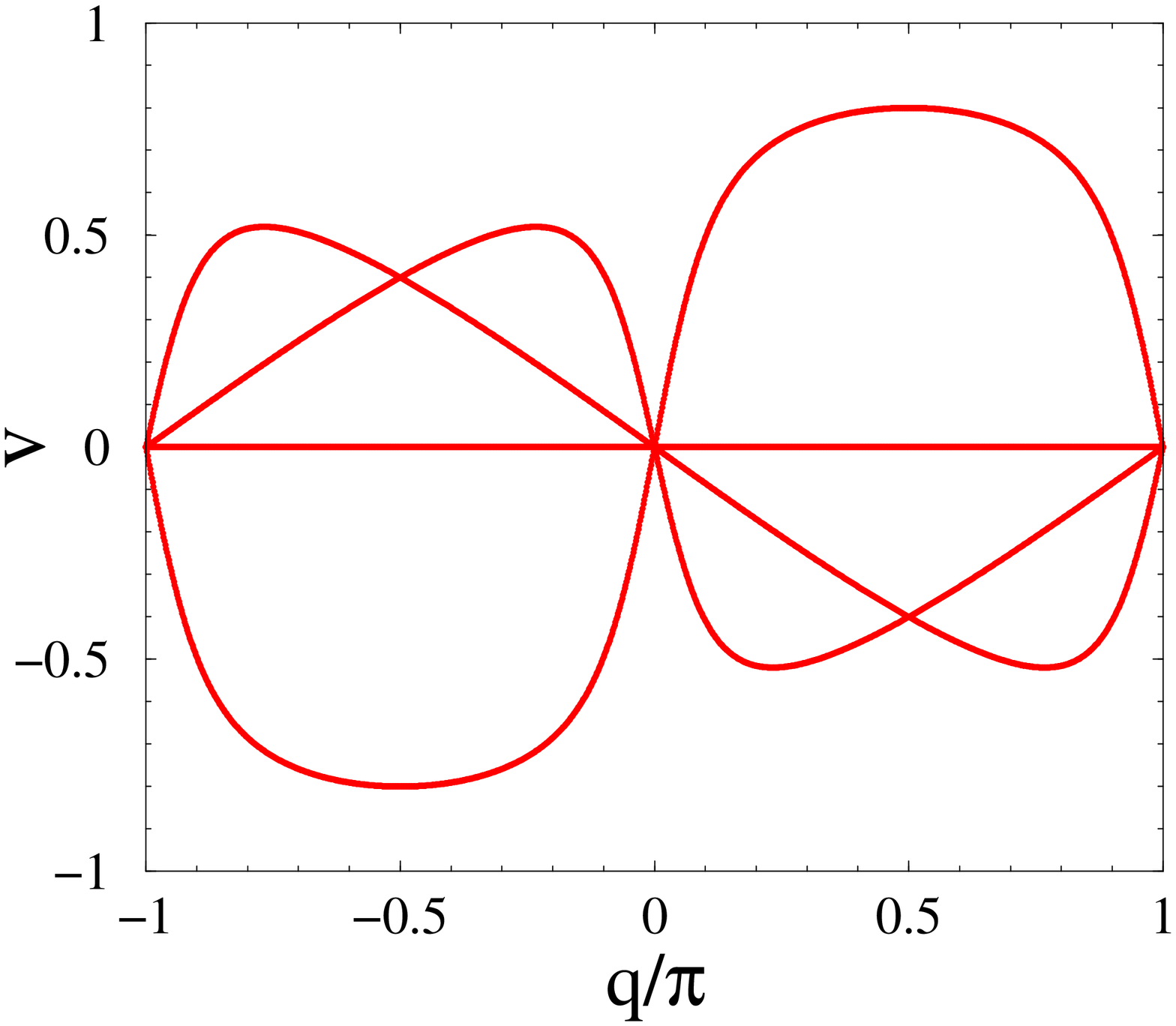}
\caption{\small
Left: energy spectrum of the $N=3$ centipede against $q/\pi$.
Right: associated group velocities.}
\label{wv3}
\end{center}
\end{figure}

The wavefunction amplitudes obey the equations
\beqa
\ii\,\frac{\d\p_n^{00}(t)}{\d t} &=& \p_{n-1}^{10}(t)+\p_n^{01}(t),
\nonumber\\
\ii\,\frac{\d\p_n^{01}(t)}{\d t} &=& \p_{n-1}^{11}(t)+\p_n^{00}(t)+\p_n^{10}(t),
\nonumber\\
\ii\,\frac{\d\p_n^{10}(t)}{\d t} &=& \p_n^{01}(t)+\p_n^{11}(t)+\p_{n+1}^{00}(t),
\nonumber\\
\ii\,\frac{\d\p_n^{11}(t)}{\d t} &=& \p_n^{10}(t)+\p_{n+1}^{01}(t).
\label{am3}
\eeqa
The corresponding Hamiltonian reads
\beq
\H=\pmatrix{0&1&\e^{\ii q}&0\cr
1&0&1&\e^{\ii q}\cr
\e^{-\ii q}&1&0&1\cr
0&\e^{-\ii q}&1&0}.
\label{ham3}
\eeq
The associated characteristic equation is
\beq
\o(\o^3-5\o-4\cos q)=0.
\label{w3}
\eeq
Figure~\ref{wv3} shows plots of the energy spectrum (left)
and of the group velocities (right) against $q/\pi$.
The maximal velocities $\pm V\ind{3}$, with
\beq
V\ind{3}=\frac{4}{5},
\label{v3}
\eeq
are respectively reached for $q=\pm\pi/2$.
The group velocity also exhibits a flat (i.e., non-dispersive) band,
as well as four non-trivial stationary points obeying~(\ref{dvdq}).
Differentiating twice the characteristic equation~(\ref{w3}),
we obtain after some algebra the stationary velocities $\pm V\ind{3,1}$, with
\beq
V\ind{3,1}=\frac{\sqrt{5-17^{1/3}}}{3}=0.519\,478\dots
\eeq

\begin{figure}[!ht]
\begin{center}
\includegraphics[angle=-90,width=.45\linewidth]{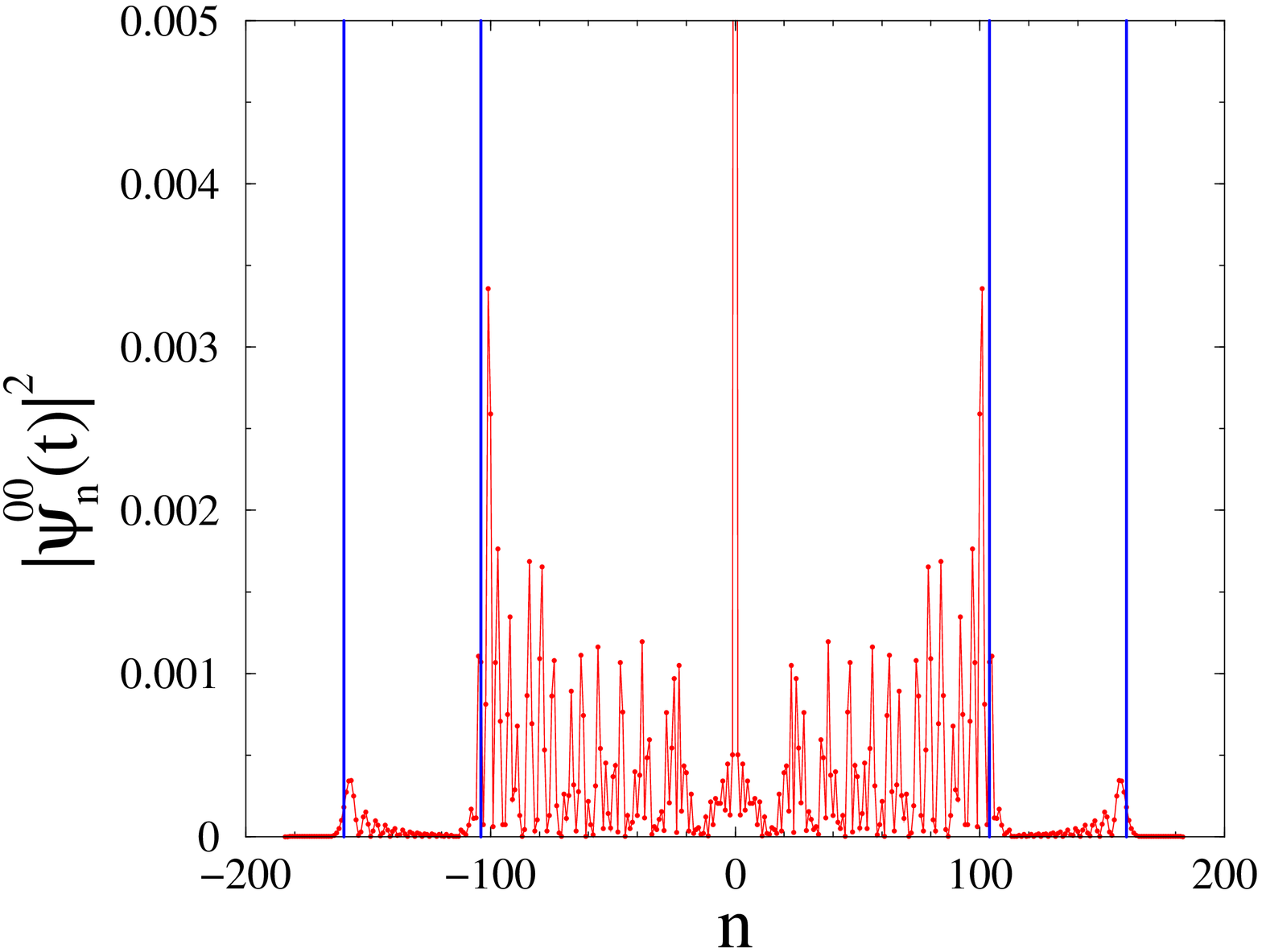}
\hskip 6pt
\includegraphics[angle=-90,width=.45\linewidth]{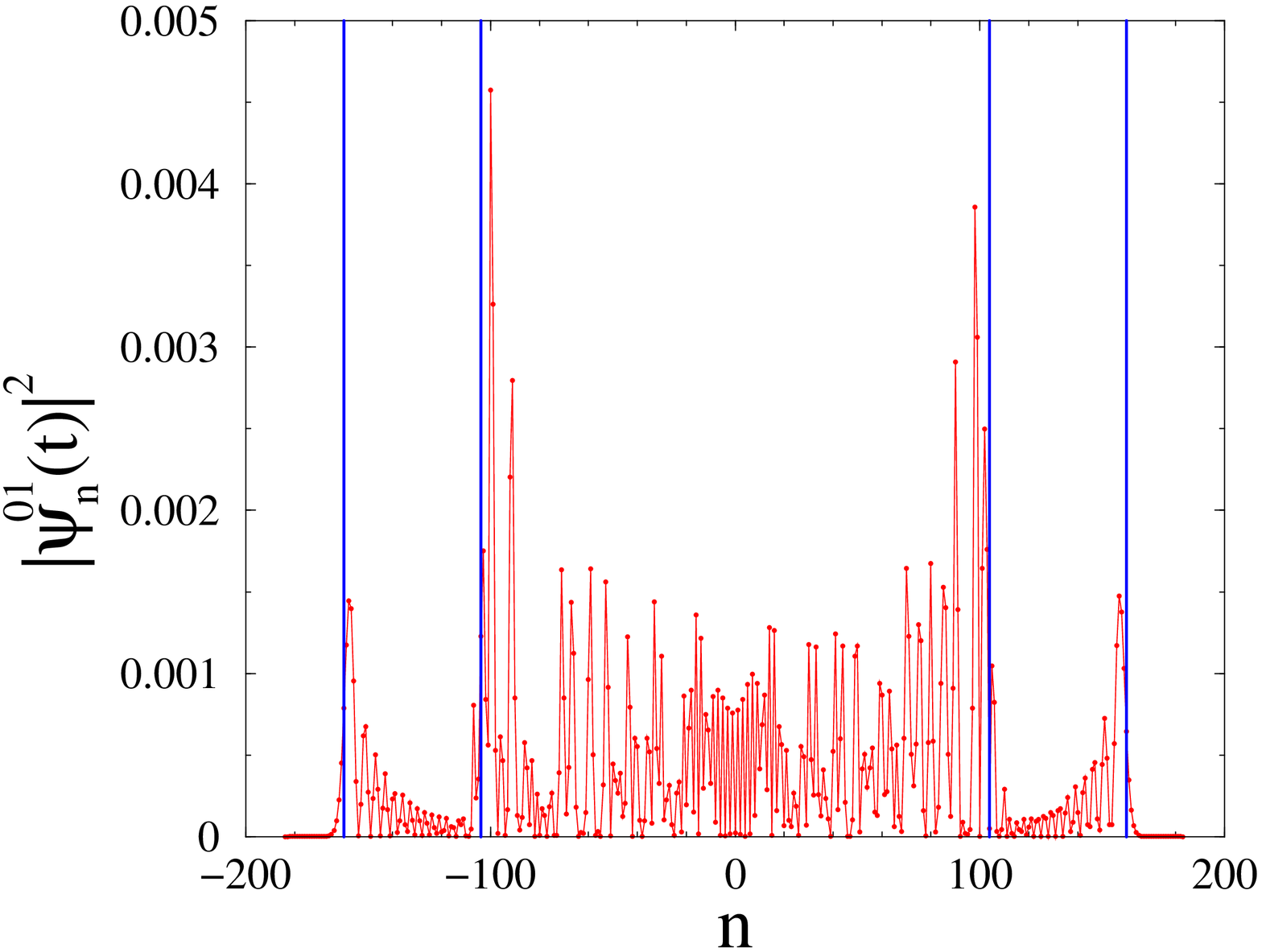}
\caption{\small
Two probability profiles at time $t=200$ for the $N=3$ centipede launched
in its most compact state near the origin ($\p_0^{00}(0)=1$).
Left: $\abs{\p_n^{00}(t)}^2$ exhibits a central peak (not to scale).
Right: $\abs{\p_n^{01}(t)}^2$ does not.
Vertical blue lines: nominal positions of the ballistic fronts at
$\pm V\ind{3}t$ and $\pm V\ind{3,1}t$.}
\label{p3}
\end{center}
\end{figure}

These results show that the wavefunction of the 3-fermion centipede
generically exhibit five ballistic peaks in the center-of-mass coordinate:
two extremal ones at $n\approx\pm V\ind{3}t$,
two internal ones at $n\approx\pm V\ind{3,1} t$,
and possibly a central one at the origin, corresponding to the flat band.
These predictions are illustrated in figure~\ref{p3},
showing plots of the probability profiles $\abs{\p_n^{00}(t)}^2$ (left)
and $\abs{\p_n^{01}(t)}^2$ (right) at time $t=200$ against~$n$
($n$ serves as a proxy for the center-of-mass coordinate $x_\cm$)
for the $N=3$ centipede launched at $t=0$ at sites 0, 1 and 2,
i.e., with a single non-zero amplitude $\p_0^{00}(0)=1$.
The first profile exhibits a central peak, whereas the second one does not.

\begin{figure}[!ht]
\begin{center}
\includegraphics[angle=-90,width=.45\linewidth]{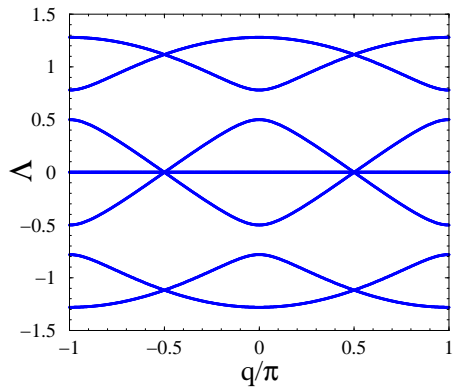}
\caption{\small
Quasiparticle spectrum of the $N=3$ centipede against~$q/\pi$.}
\label{h3}
\end{center}
\end{figure}

Equation~(\ref{chni}) yields a cubic equation for the quasiparticle eigenvalues:
\beq
4\La^3-5\La\pm2\cos q=0.
\eeq
A comparison with~(\ref{w3})
demonstrates that the correspondence~(\ref{ohl}) goes as follows for $N=3$:
the energies $\o$ are twice as large as (some of) the quasiparticle eigenvalues~$\La$.
Figure~\ref{h3} shows the quasiparticle spectrum against~$q/\pi$.

\subsection*{$\bullet$ $N=4$}

\begin{figure}[!ht]
\begin{center}
\includegraphics[angle=-90,width=.45\linewidth]{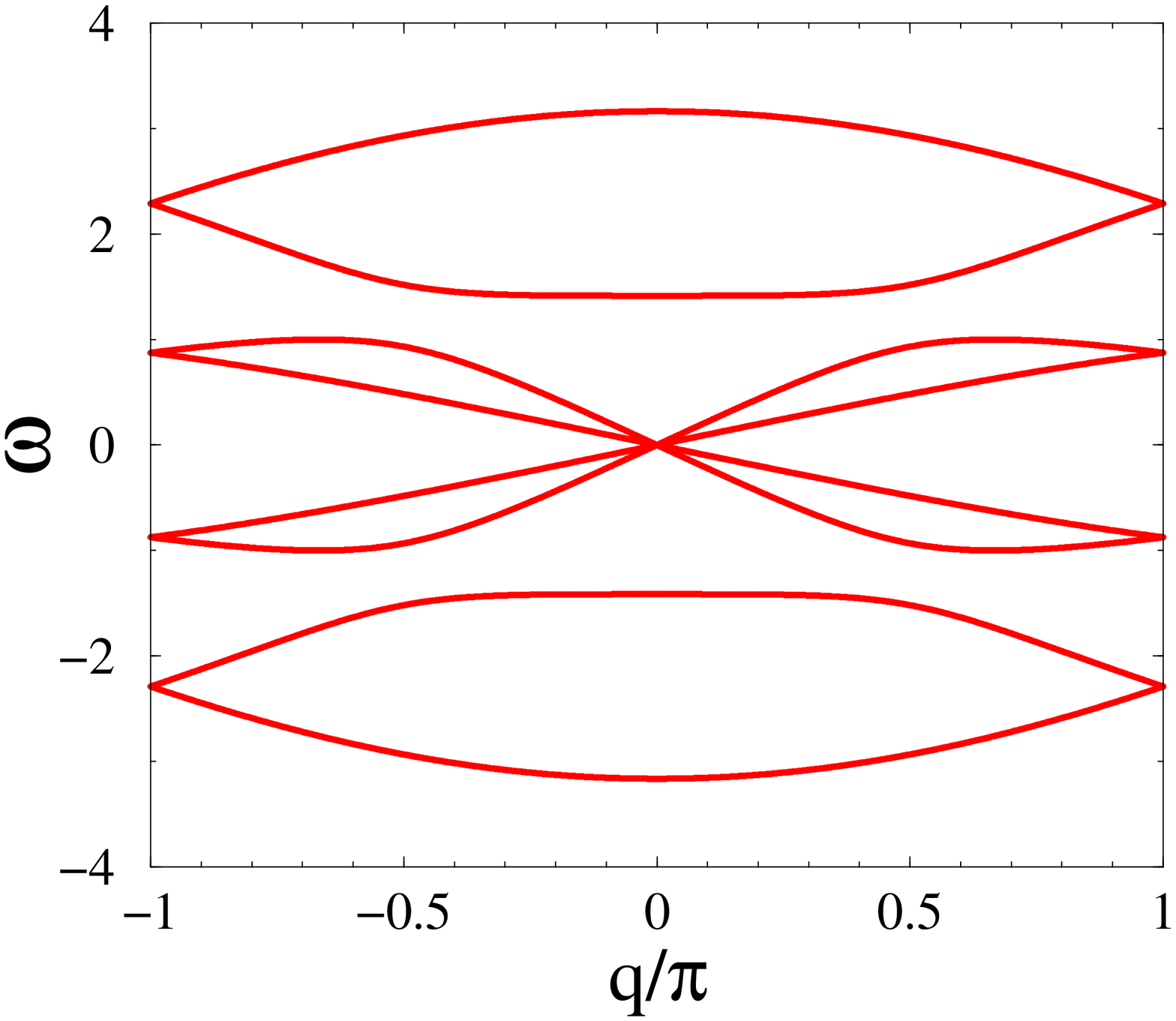}
\includegraphics[angle=-90,width=.45\linewidth]{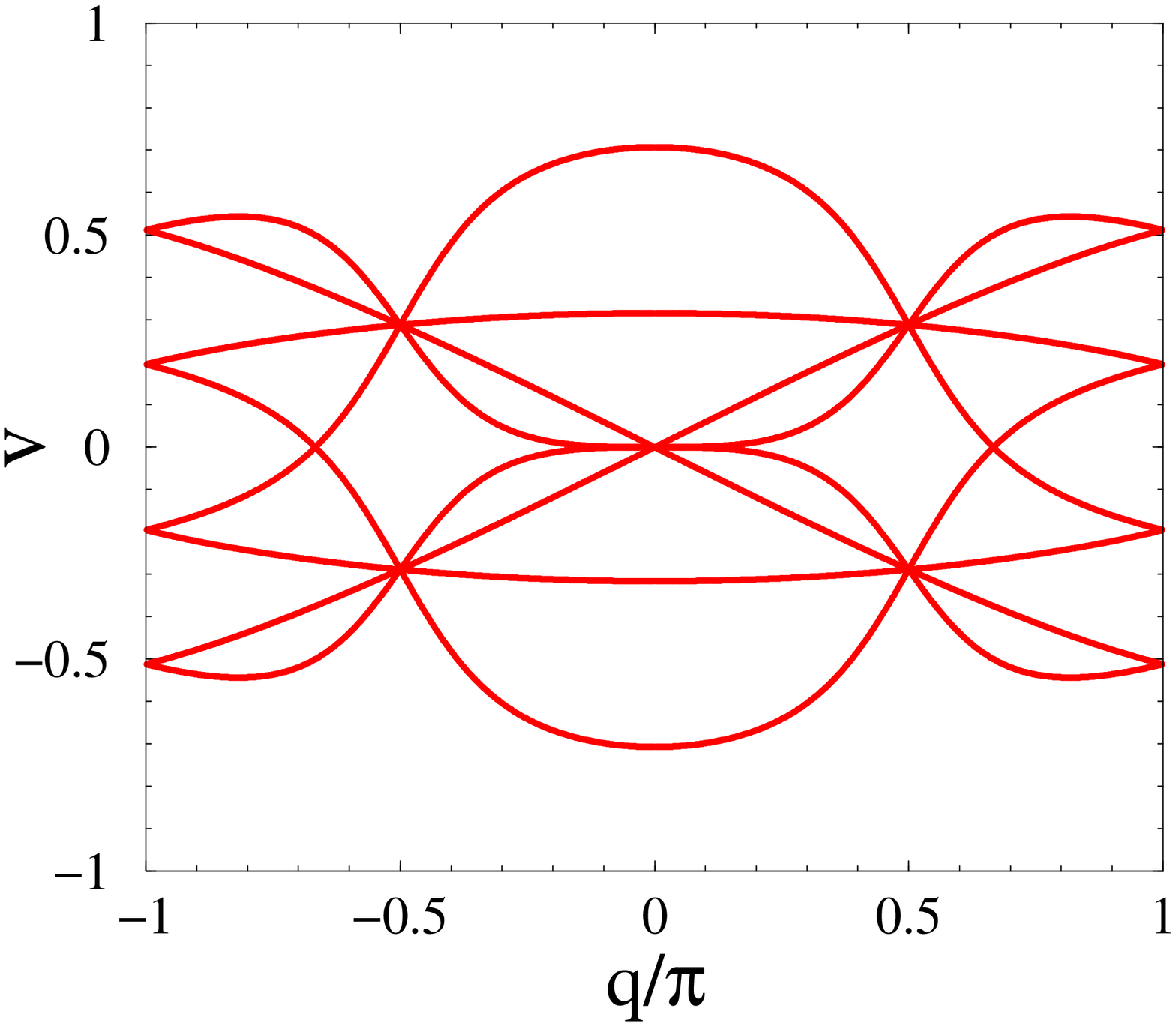}
\caption{\small
Left: energy spectrum of the $N=4$ centipede against $q/\pi$.
Right: associated group velocities.}
\label{wv4}
\end{center}
\end{figure}

\begin{figure}[!ht]
\begin{center}
\includegraphics[angle=-90,width=.45\linewidth]{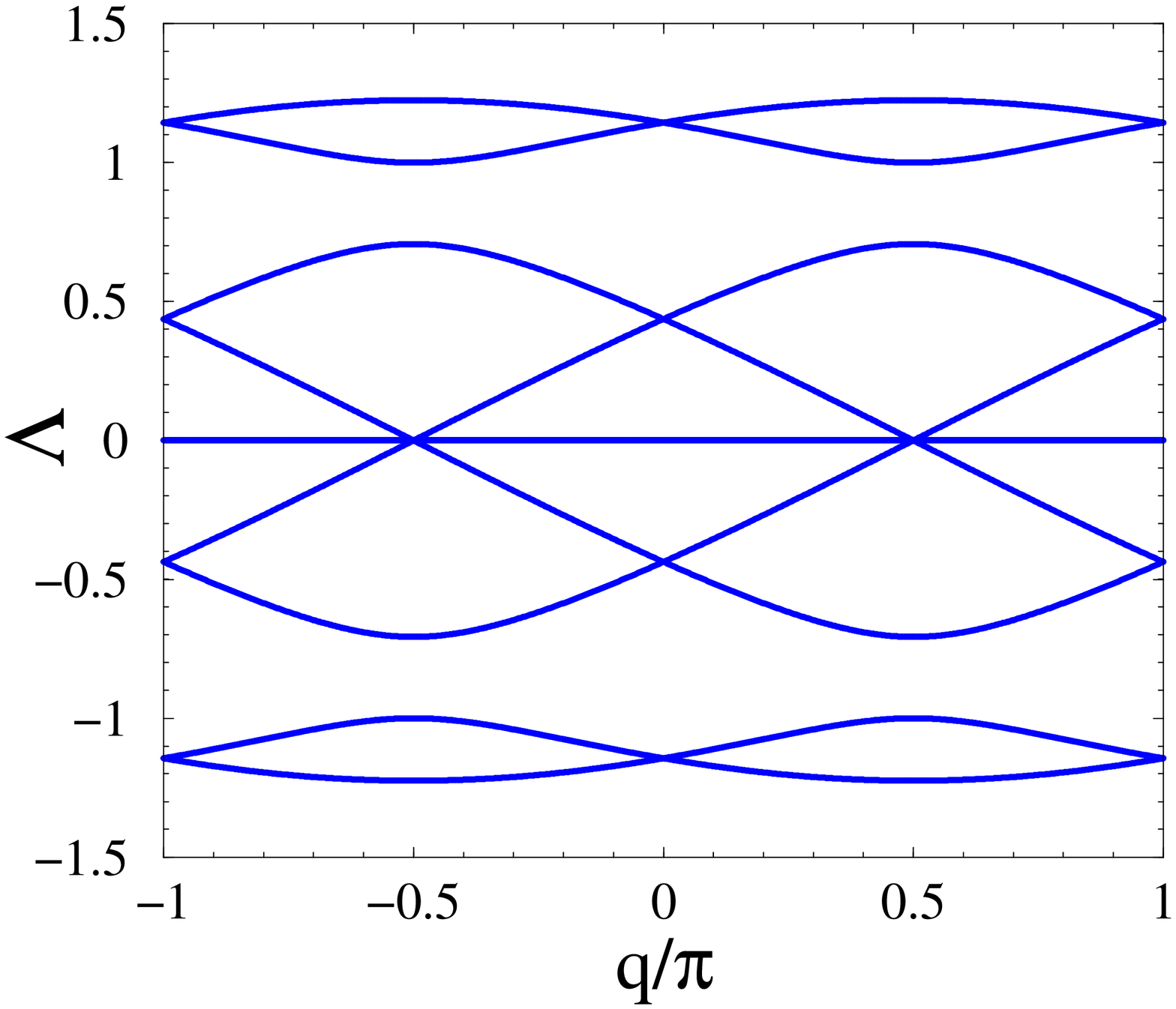}
\caption{\small
Quasiparticle spectrum of the $N=4$ centipede against~$q/\pi$.}
\label{h4}
\end{center}
\end{figure}

The Hamiltonian reads
\beq
\H=\pmatrix{
0&1&0&0&\e^{\ii q}&0&0&0\cr
1&0&1&0&0&\e^{\ii q}&0&0\cr
0&1&0&1&1&0&\e^{\ii q}&0\cr
0&0&1&0&0&1&0&\e^{\ii q}\cr
\e^{-\ii q}&0&1&0&0&1&0&0\cr
0&\e^{-\ii q}&0&1&1&0&1&0\cr
0&0&\e^{-\ii q}&0&0&1&0&1\cr
0&0&0&\e^{-\ii q}&0&0&1&0}.
\label{ham4}
\eeq
The associated characteristic equation is
\beq
\o^8-12\o^6+4(8-3\cos q)\o^4-24(1-\cos q)\o^2+4(1-\cos q)^2=0.
\eeq
Figure~\ref{wv4} shows plots of the energy spectrum (left)
and of the group velocities (right) against $q/\pi$.
The maximal velocities $\pm V\ind{4}$, with
\beq
V\ind{4}=\frac{1}{\sqrt{2}},
\label{v4}
\eeq
are reached for $q=0$.
Figure~\ref{h4} shows the quasiparticle spectrum against~$q/\pi$.
This is the first case where the correspondence~(\ref{ohl}) exhibits its generic nature,
in the sense that it involves non-trivial linear combinations.

\subsection*{$\bullet$ $N=5$}

\begin{figure}[!ht]
\begin{center}
\includegraphics[angle=-90,width=.45\linewidth]{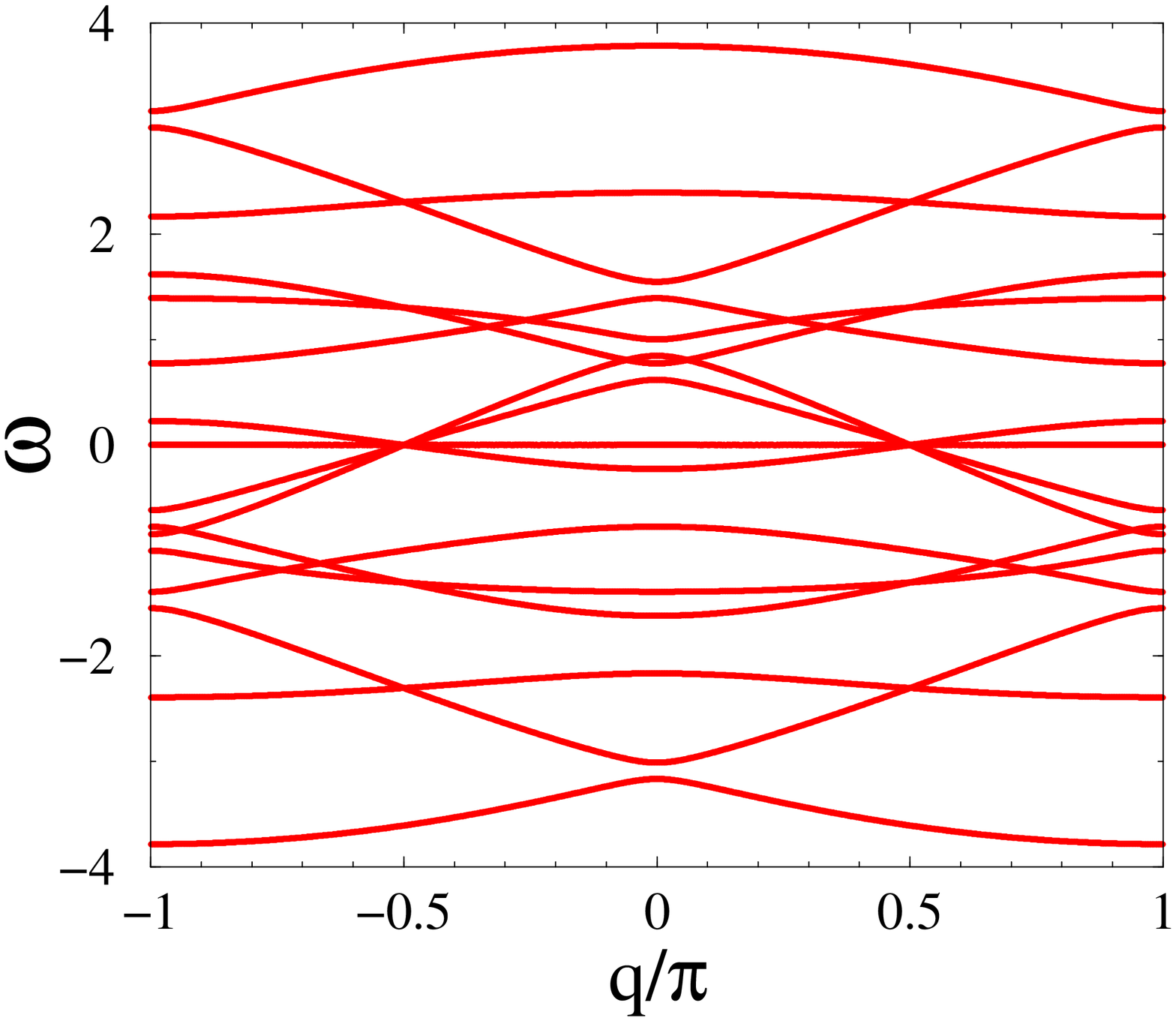}
\includegraphics[angle=-90,width=.45\linewidth]{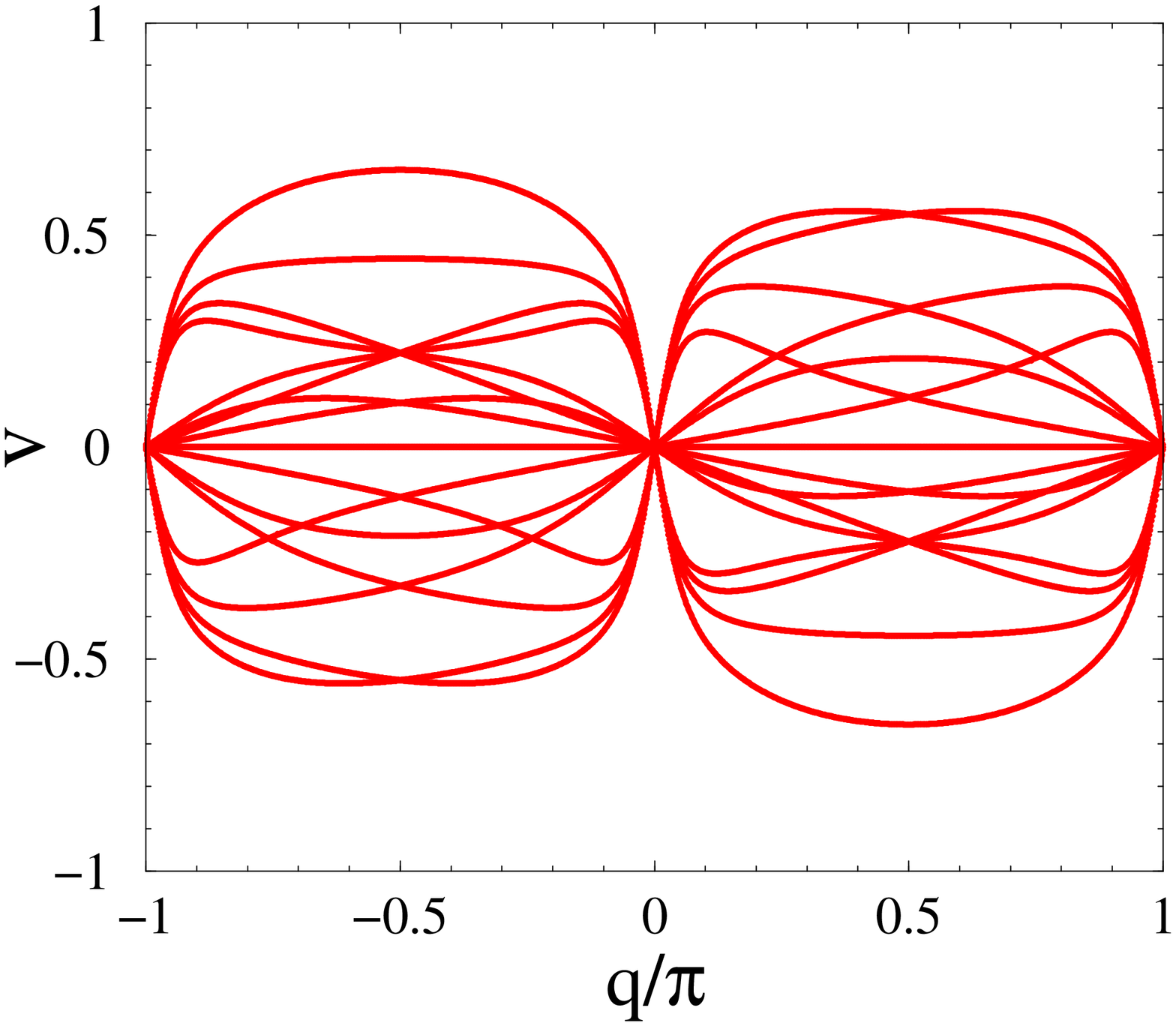}
\caption{\small
Left: energy spectrum of the $N=5$ centipede against $q/\pi$.
Right: associated group velocities.}
\label{wv5}
\end{center}
\end{figure}

\begin{figure}[!ht]
\begin{center}
\includegraphics[angle=-90,width=.45\linewidth]{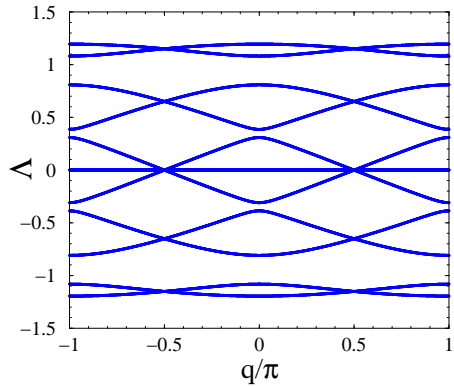}
\caption{\small
Quasiparticle spectrum of the $N=5$ centipede against~$q/\pi$.}
\label{h5}
\end{center}
\end{figure}

We shall not write down the $16\times 16$ Hamiltonian matrix $\H$ explicitly.
The associated characteristic equation is
\beq
\o(\o^5-7\o^3+9\o-4\cos q)(A(\o)-B(\o)\cos q-16\cos^2q)=0,
\label{w5}
\eeq
with
\beqa
&&A(\o)=\o^2(\o^2-1)(\o^2-13)(\o^4-7\o^2+9),
\nonumber\\
&&B(\o)=4\o(11\o^4-28\o^2+13).
\eeqa
Figure~\ref{wv5} shows plots of the energy spectrum (left)
and of the group velocities (right) against $q/\pi$.
The maximal velocities $\pm V\ind{5}$, with
\beq
V\ind{5}=\frac{26+14\sqrt{13}}{117}=0.653\,655\dots,
\label{v5}
\eeq
are respectively reached for $q=\mp\pi/2$.
Figure~\ref{h5} shows the quasiparticle spectrum against~$q/\pi$.

The above results illustrate the general feature that the complexity
of the energy spectrum of the centipede
grows very fast as the fermion number $N$ is increased.
In particular the number of stationary values of the velocity satisfying~(\ref{dvdq}),
which are responsible for the occurrence of internal ballistic fronts,
grows very rapidly with $N$.
The quasiparticle spectrum however remains regular and simple,
as the number of quasiparticle eigenvalues only grows linearly with $N$.

\section{Maximal spreading velocity for arbitrary $N$}
\label{vmax}

The aim of this section is to obtain an exact expression
of the maximal velocity $V\ind{N}$ of the centipede for an arbitrary fermion number $N$.
This quantity, defined in~(\ref{vmaxdef}),
characterizes the ballistic spreading of the two extremal fronts of the wavefunction
in the center-of-mass coordinate.
We shall also derive the exact value of the limit~$V\ind{\infty}$.

The explicit results given in section~\ref{explicit} for the first few values of~$N$
suggest the following pattern.
The maximal velocity $V\ind{N}$ is reached
for the values of the center-of-mass momentum $q$ such that the right-hand side
of the characteristic equations~(\ref{chnp}),~(\ref{chni}) vanishes,
i.e., $q=0$ or $\pm\pi$ when $N$ is even, whereas $q=\pm\pi/2$ when $N$ is odd.
In these situations, all the quasiparticle eigenvalues~$\La_k$ are twofold degenerate.
The simultaneous linear lifting of these degeneracies yields, by means of~(\ref{ohl}),
the largest possible value of the group velocity.

Let us consider for definiteness the case where $N$ is even, and set
\beq
\frac{\sin((N+1)p)-3\sin((N-1)p)}{\sin p}=P_N(\La),
\label{pdef}
\eeq
with $\La=\cos p$ (see~(\ref{lp}),~(\ref{val:La})).
The function $P_N(\La)$ thus defined is a polynomial with degree $N$,
which can be expressed as a linear combination of two Chebyshev polynomials
of the second kind~\cite{GR}:
\beq
P_N(\La)=U_N(\La)-3U_{N-2}(\La).
\eeq
The polynomials $P_N$ obey the recursion
\beq
P_{N+1}(\La)=2\La P_N(\La)-P_{N-1}(\La).
\eeq
We have $P_0(\La)=4$,
$P_1(\La)=2\La$,
$P_2(\La)=4(\La^2-1)$,
$P_3(\La)=2\La(4\La^2-5)$,
and so on.
The characteristic equation~(\ref{chnp}) can thus be recast as
\beq
P_N(\La)=\pm4\sin q.
\eeq
For $q=0$ or $q=\pi$, the right-hand side vanishes.
The doubly degenerate quasiparticle eigenvalues therefore coincide
with the $N$ roots $\La_k$ of the polynomial~$P_N$.
The lifting of these twofold degeneracies in the vicinity of $q=0$ or $q=\pi$
is described by the~slopes
\beq
\frac{\d\La_k}{\d q}=\pm\frac{4}{P_N'(\La_k)}.
\eeq
Using~(\ref{ohl}), taking care about avoiding multiple counting,
we obtain the expression
\beq
V\ind{N}=\sum_{k=1}^N\frac{4}{\abs{P_N'(\La_k)}}
\label{vnres}
\eeq
for the maximal spreading velocity.
The case where $N$ is odd can be dealt with in a similar way
and yields the same expression.

The general formula~(\ref{vnres})
allows us to recover~(\ref{v2}),~(\ref{v3}),~(\ref{v4}),~(\ref{v5}), i.e.,
\beqa
V\ind{2}=1,\quad
V\ind{3}=\frac{4}{5}=0.8,
\nonumber\\
V\ind{4}=\frac{1}{\sqrt{2}}=0.707\,106\dots,\quad
V\ind{5}=\frac{26+14\sqrt{13}}{117}=0.653\,655\dots
\eeqa
and to predict that
\beq
V\ind{6}=0.620\,924\dots,\quad
V\ind{7}=0.600\,722\dots
\eeq
are the largest roots of the polynomial equations
$229V^4-78V^2-8V+1=0$ and $79\,937V^3-49\,192V^2-3\,664V+2\,624=0$,
with respective degrees 4 and 3.
More generally, the maximal velocity $V\ind{N}$ is an algebraic number
whose degree $d_N$ grows exponentially fast with $N$.
Let us skip details and give the following result:
\beqa
N=2m\hbox{ even:}\hfill & d_{2m}=2^{m-1},
\nonumber\\
N=2m+1\hbox{ odd:}\qquad & d_{2m+1}={m\choose \Int(m/2)},
\eeqa
where $\Int(.)$ denotes the integer part.

As the fermion number $N$ increases,
the velocities $V\ind{N}$ converge to a finite limit~$V\ind{\infty}$,
which can be obtained as follows.
For $N\ge3$, $N-2$ roots $\La_k$ of the polynomial~$P_N$ obey $\abs{\La_k}<1$.
They correspond to real momenta $p_k$, such that
\beq
Np_k=k\pi+\theta_k\qquad(k=1,\dots,N-2),
\label{npk}
\eeq
with $\abs{\theta_k}\le\pi/2$ and
\beq
\tan\theta_k=2\tan p_k.
\label{tata}
\eeq
The last two roots satisfy $\abs{\La}>1$.
They correspond to evanescent modes with complex momenta
$p=\ii\zeta$ and $p=\pi+\ii\zeta$,
such that $\La=\pm\cosh\zeta$, with $\tanh(N\zeta)=2\tanh\zeta$.
In the large-$N$ limit, we have $\tanh\zeta\to1/2$,
and so $\zeta\to(\ln 3)/2$ (see~(\ref{zeta})) and $\La\to\pm2/\sqrt3$.

By differentiating~(\ref{pdef}), using~(\ref{npk}) and~(\ref{tata}),
we obtain the following estimate
\beq
\abs{P_N'(\La_k)}\approx\frac{2N}{\sin^2p_k}
\underbrace{\left(\cos\theta_k\cos p_k+2\sin\theta_k\sin p_k\right)}
_{\textstyle{\sqrt{1+3\sin^2p_k}}}
\eeq
for large $N$ and real momenta $p_k$.
Finally, inserting the above expression into~(\ref{vnres}),
and replacing the sum by an integral, we obtain
\beqa
V\ind{\infty}
&=&\frac{4}{\pi}\int_0^{\pi/2}\frac{\sin^2p\,\d p}{\sqrt{1+3\sin^2p}}
\nonumber\\
&=&\frac{2}{3\pi}\left(4\EE(\frat{\sqrt{3}}{2})-\KK(\frat{\sqrt{3}}{2})\right)
=0.570\,349\,449\dots,
\label{vinfres}
\eeqa
where $\EE$ and $\KK$ are the complete elliptic integrals~\cite{GR}.

\begin{figure}[!ht]
\begin{center}
\includegraphics[angle=-90,width=.45\linewidth]{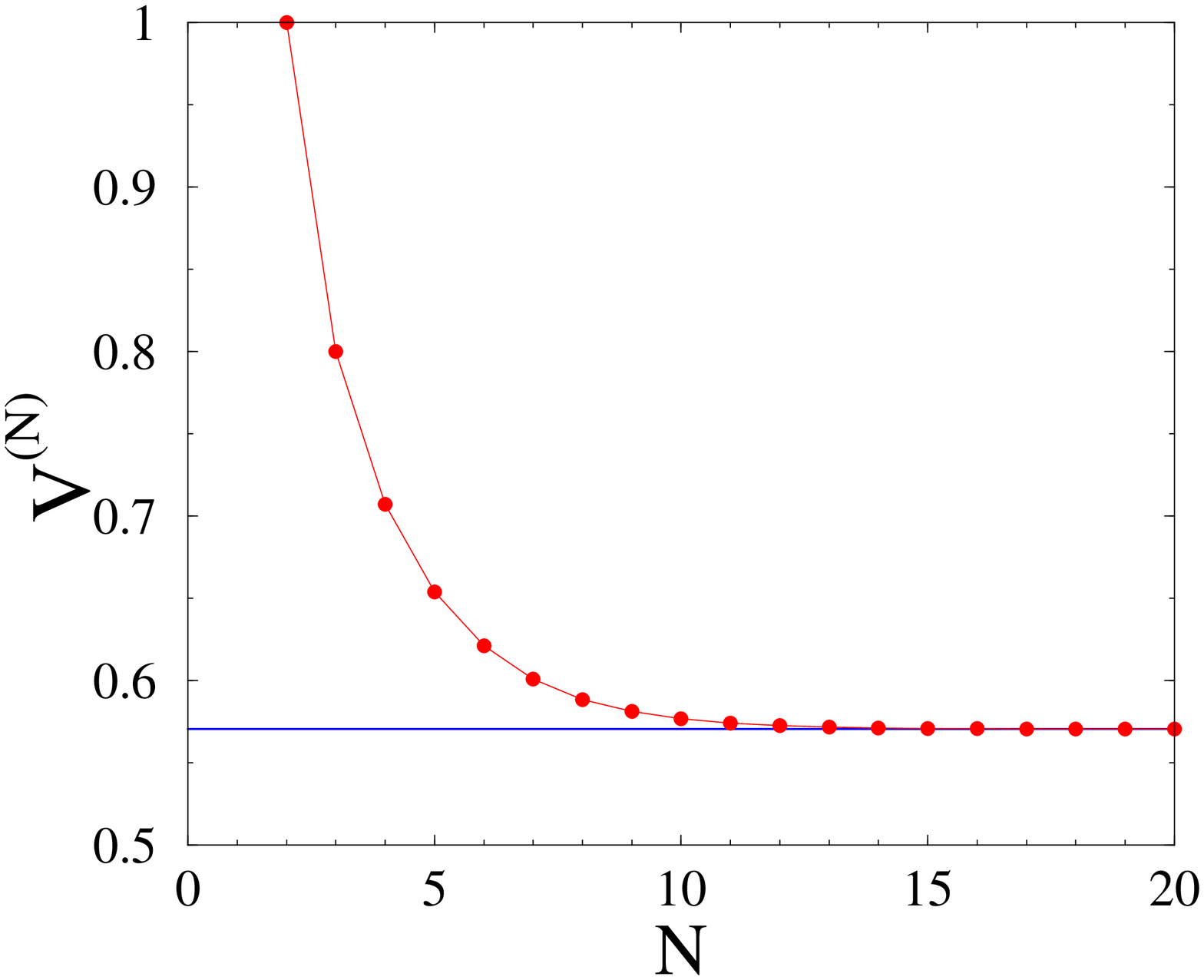}
\includegraphics[angle=-90,width=.45\linewidth]{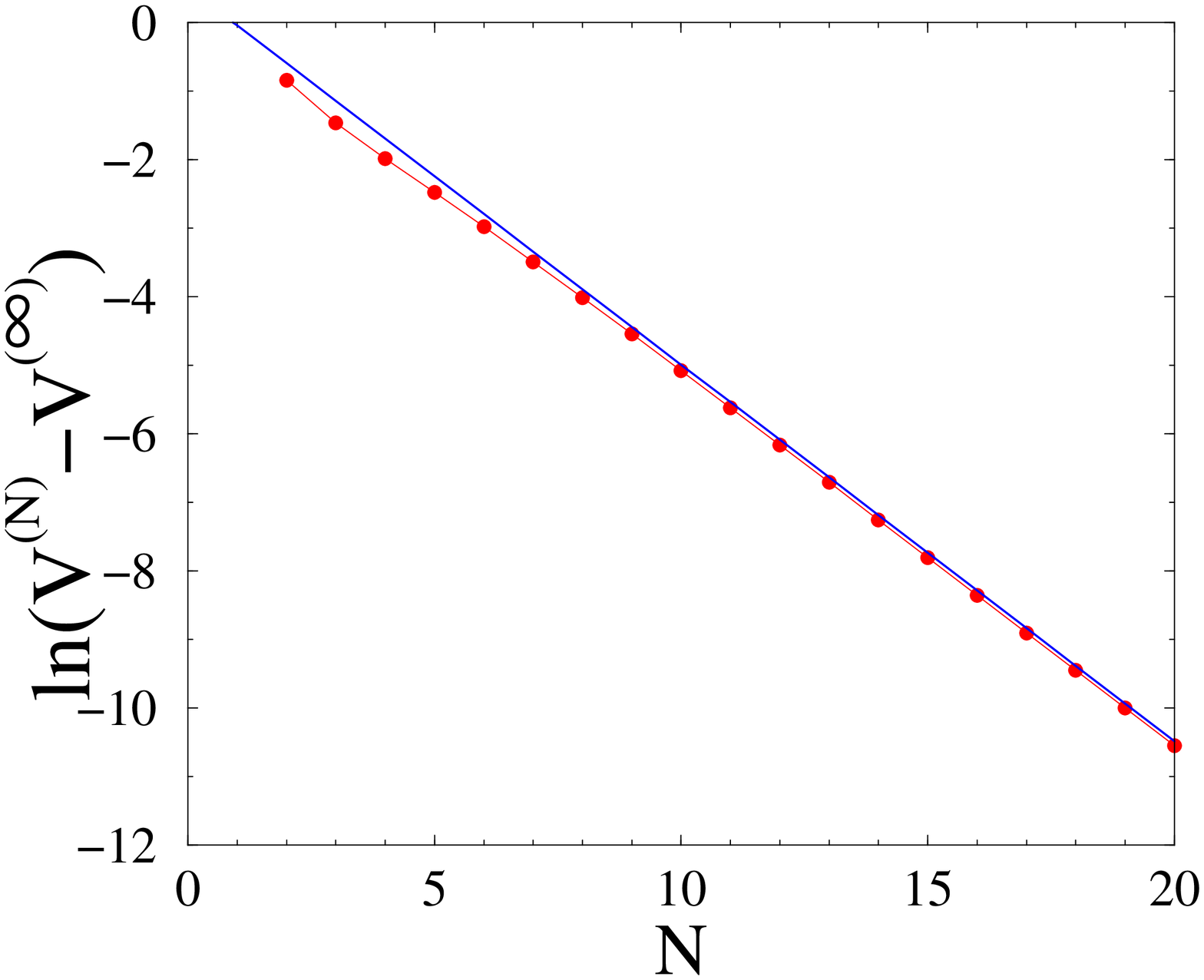}
\caption{\small
Left: plot of the extremal velocity $V\ind{N}$ against the fermion number $N$.
Blue horizontal line: limit $V\ind{\infty}$ (see~(\ref{vinfres})).
Right: logarithmic plot of difference $V\ind{N}-V\ind{\infty}$ against $N$.
The blue straight line has slope $-\zeta$.}
\label{vgd}
\end{center}
\end{figure}

Figure~\ref{vgd} illustrates the above results.
The left panel shows a plot of $V\ind{N}$ against the fermion number $N$.
The limit $V\ind{\infty}$ (see~(\ref{vinfres})) is shown as a blue line.
The right panel shows a logarithmic plot of the difference $V\ind{N}-V\ind{\infty}$
against $N$.
The data points are observed to become extremely close
to the straight line with slope $-\zeta$, where
\beq
\zeta=\frac{\ln 3}{2}=0.549\,306\dots
\label{zeta}
\eeq
is the inverse penetration length of the evanescent modes.
This clearly demonstrates that the velocities converge to their limit
as $V\ind{N}-V\ind{\infty}\sim\e^{-N\zeta}$,
i.e., exponentially fast in $N$.

\section{Discussion}
\label{discussion}

We have investigated a quantum centipede
made of $N$ fermionic quantum walkers on the one-dimensional lattice
interacting by the hard constraint
that the distance between two successive fermions is either one or two lattice spacings.
Besides the number $N$ of fermions, the model is entirely parameter-free.

As in our previous work~\cite{KLM},
the main emphasis has been put on the ballistic spreading
of the wavefunction of the centipede in its center-of-mass coordinate.
For a generic initial state located in the vicinity of the origin,
the distribution profile of the velocity $v=n/t$ of the center of mass
generically exhibits two extremal ballistic fronts at $\pm V\ind{N}$,
as well as internal ballistic fronts, whose number grows rapidly
with the number $N$ of fermions.

The energy spectrum of the centipede
and the corresponding velocity dispersion curve
have been analyzed by direct means for the first few values of $N$,
whereas some analytical results have been derived for arbitrary $N$
by exploiting a mapping of the problem onto a free-fermion system.
We have thus obtained the expression~(\ref{vnres})
of the maximal spreading velocity $V\ind{N}$,
and the non-trivial result~(\ref{vinfres}) for the limit~$V\ind{\infty}$.

It is interesting to put the present findings in perspective
with the results of~\cite{Tibor} on the diffusive dynamics
of $N$-legged molecules dubbed polypeds and spiders~\cite{exp}.
The classical analogue of the present situation is that of symmetric molecular centipedes,
whose diffusion coefficient is given (for all $N\ge2$) by
\beq
D\ind{N}=\frac{1}{4(N-1)}.
\eeq
There is a stark contrast between the fall-off of the diffusion coefficient
$D\ind{N}$ for large~$N$ in the classical case
and the convergence of the spreading velocity to a finite limit~$V\ind{\infty}$
in the quantum case.
This is yet another manifestation of the qualitatively different dynamical behavior
of classical and quantum walkers.

Some of the properties of the quantum centipede
depend on the parity of the fermion number $N$.
The symmetries of the energy spectrum
ensure the existence of a flat (i.e., non-dispersive) band when $N$ is odd.
As a consequence, the wavefunction of the centipede may exhibit
a central peak near the origin for odd $N$.
(This is illustrated in figure~\ref{p3} for $N=3$.)
The occurrence of a central peak has been underlined in other types of quantum walks.
For a single discrete-time walker equipped with a three-dimensional quantum coin,
a localization phenomenon has been put forward,
in the sense that a finite fraction of the probability stays forever
in the vicinity of the particle's starting point~\cite{iks,sbj}.
From a different perspective, parity effects are also known to affect
transport properties of some quasi-one-dimensional systems.
Disordered strips made of $N$ coupled channels with purely off-diagonal disorder
are known to exhibit conventional Anderson localization for even~$N$,
albeit unconventional localization properties for odd $N$,
with a subexponential scaling of the conductance
at the band center~\cite{str1,str2,str3,str4}.

In this work we have demonstrated that the simplest fermionic quantum centipede,
with maximal separation $\ell=2$ between neighboring particles,
is tractable by analytical means.
It would be interesting to investigate fermionic or bosonic quantum centipedes
with larger maximal separations as well.
Classical centipedes with $\ell\ge3$ however lead to extremely complicated results,
so that the general case seems intractable~\cite{Tibor}.
Another variant that has been studied in the classical case
is a centipede whose total length never exceeds some given length~$L$~\cite{Tibor}.
Its quantum analogue also appears to be interesting.
Finally, it might also be worth considering bound states of $N$ quantum walkers,
either fermionic or bosonic,
on higher-dimensional lattices with various kinds of hard-bound constraints.

\appendix

\section{Derivation of the characteristic equations~(\ref{chnp}),~(\ref{chni})}
\label{app}

In this appendix we provide a detailed characterization of the quasiparticle operators
defined in~(\ref{aadag}) and a derivation
of the characteristic equations~(\ref{chnp}),~(\ref{chni}).

Our starting point is the quadratic identity
\beq
\lbrack \tau_{j}^\mu \tau_{j+1}^\nu, \tau_{k}^\lambda \rbrack
= 2\left( \delta_{j+1,k}\delta^{\nu\lambda} \tau_{j}^\mu
- \delta_{jk}\delta^{\mu\lambda} \tau_{j+1}^\nu\right),
\eeq
where $\mu,\nu,\lambda=x,y$,
that follows from the Clifford algebra~(\ref{Clifford}).
We can then write explicitly the commutation relation
$\lbrack \H_\lg, a_k \rbrack = - 2 \La_k a_k $ (see~(\ref{eq:Heisenberg})) as follows
\beqa
&\;\ii \cos{q} \left( x_1 \, \tau_{0}^y - y_0 \, \tau_{1}^x \right)
+ \ii \sin{q} \left( - y_0 \, \tau_{1}^y + y_1 \, \tau_{0}^y \right)
\nonumber \\
+&\;\frac{\ii}{2} \sum_{j=1}^{N-2}
\left( x_{j+1} \, \tau_{j}^y + x_{j}\, \tau_{j+1}^y
- y_{j}\,\tau_{j+1}^x - y_{j+1}\, \tau_{j}^x \right)
\nonumber \\
+&\;\ii \left( x_N \, \tau_{N-1}^y - y_{N-1} \tau_N^x \right)
= \La \sum_{j=0}^N \left( x_{j} \, \tau_{j}^x + y_{j}\, \tau_{j}^y \right).
\eeqa
(The subscript $k=0,\dots,N$ has been omitted for ease of reading.)
After identifying the coefficients of each $\tau_{j}^{x,y}$, we obtain the following
system of linear equations for the coefficients $x_j$ and $y_j$:

\smallskip

\noindent
For $j=0$:
\beq
x_0=0,\quad -\ii \La y_0 = x_1 \cos{q} + y_1 \sin{q}.
\label{eq:j0}
\eeq
For $j=1$:
\beq
2 \ii \La x_1 = 2 y_0 \cos q + y_2,\quad
- 2 \ii \La y_1 = x_2 - 2 y_0 \sin{q}.
\label{eq:j1}
\eeq
For $j=2,\dots,N-2$:
\beq
2 \ii \La x_j = \, y_{j-1} + y_{j+1},\quad
-2 \ii \La y_j = \, x_{j-1} + x_{j+1}.
\label{eq:bulk}
\eeq
For $j=N-1$:
\beq
2 \ii \La x_{N-1} = y_{N-2},\quad
-2 \ii \La y_{N-1} = x_{N-2} + 2 x_N.
\label{eq:jL}
\eeq
For $j=N$:
\beq
\ii \La x_N = y_{N-1},\quad y_N = 0.
\label{eq:jLplus1}
\eeq
The above equations can be interpreted as an eigenvalue problem for the vector
$ (x_{0}, y_{0}, x_{1}, y_{1}, \dots, x_N, y_N)$ of length $2(N+1)$.
The quasiparticle eigenvalues $\La$ are obtained as follows.
The bulk equations~(\ref{eq:bulk}) yield the recursion
\beq
4 \La^2 x_j = x_{j-2} + 2 x_j + x_{j+2}\qquad(j=2,\dots,N-2).
\eeq
The corresponding characteristic equation is bi-quadratic:
$ 4 \La^2 = r^{-2} + 2 + r^{2}=(r^{-1}+r)^2$.
The four values $r = \pm \e^{ \pm \ii p}$ lead to the dispersion relation
\beq
\La = \cos{p}.
\label{val:La}
\eeq
Using~(\ref{eq:bulk}) and~(\ref{val:La}), the corresponding eigenvectors read
\beqa
{\hskip 10pt}x_j&=&A\e^{\ii jp}+B\e^{-\ii jp}+(-1)^j(C\e^{\ii jp}+D\e^{-\ii jp}),
\nonumber\\
-\ii y_j&=&A\e^{\ii jp}+B\e^{-\ii jp}-(-1)^j(C\e^{\ii jp}+D\e^{-\ii jp}).
\label{sol:generic}
\eeqa
The values of $p$ are yet to be determined by the boundary conditions.
First, using~(\ref{eq:j0}) and~(\ref{eq:jLplus1}), we substitute
$y_0$ and $x_N$ in~(\ref{eq:j1}) and~(\ref{eq:jL}) to obtain
\beqa
2 \La^2 x_1 = (1+\cos2q) x_1 + \sin2q \, y_1 - \ii \La y_2,
\nonumber\\
2 \La^2 y_1 = \sin2q \, x_1 + (1-\cos2q) y_1 + \ii \La x_2,
\nonumber\\
2 \ii \La x_{N-1} = y_{N-2},
\nonumber\\
2 \La^2 y_{N-1} = \ii \La x_{N-2} + 2 y_{N-1}.
\label{xy}
\eeqa
Then, imposing that the generic forms~(\ref{sol:generic})
remain valid for $j=1$ and $j=N-1$,~(\ref{xy}) yields
\beqa
&&\e^{2\ii q}(\e^{\ii p}A+\e^{-\ii p}B)-\ii\sin p\,(C-D)=0,
\nonumber\\
&&\ii\sin p\,(A-B)-\e^{-2\ii q}(\e^{\ii p}C+\e^{-\ii p}D)=0,
\nonumber\\
&&\e^{\ii Np}A+\e^{-\ii Np}B-(-1)^N(\e^{\ii Np}C+\e^{-\ii Np}D)=0,
\nonumber\\
&&\e^{\ii Np}(\e^{\ii p}-3\e^{-\ii p})A+\e^{-\ii Np}(\e^{-\ii p}-3\e^{\ii p})B
\nonumber\\
&&{\hskip 14pt}+(-1)^N(\e^{\ii Np}(\e^{\ii p}-3\e^{-\ii p})C
+\e^{-\ii Np}(\e^{-\ii p}-3\e^{\ii p})D)=0.
\label{4par4}
\eeqa
Expressing that the $4\times4$ determinant of this system vanishes, we get
\beqa
&&\cos(2(N+1)p) - 6 \cos(2Np) + 9 \cos(2(N-1)p)
\nonumber\\
&&{\hskip 14pt}= 2 ( 1 + \cos 2p) + 8 (-1)^N \cos 2q\, ( 1 - \cos 2p).
\label{eq:caract}
\eeqa

The characteristic equation thus obtained
is a polynomial equation with degree $N+1$ in the variable $\cos 2p=2\La^2-1$.
The value $p=0$ is however not allowed.
The eigenvectors constructed as above indeed vanish identically for $p=0$.
We are thus left with $N$ pairs of opposite quasiparticle eigenvalues $\pm\La_k$.
This spectrum is to be completed by $\La=0$ with multiplicity two.
For $\e^{\ii p}=\pm\ii$ the system~(\ref{4par4})
indeed always admits the solution $A=B=C=D$, irrespective of $q$.
This elementary non-dispersive solution had been discarded
in the algebra leading to~(\ref{eq:caract}).

Inserting the above quasiparticle spectrum into~(\ref{ohl})
yields the $2^{N+1}$ (possibly degenerate)
eigenvalues $\omega_\lg$ of $\H_\lg$.
A quarter of them, corresponding to the sector $(+1,+1)$
of the boundary operators~$S_0^{x}$ and $S_N^{x}$,
coincide with the $2^{N-1}$ (again possibly degenerate) eigenvalues $\o$ of $\H$.

Finally, the characteristic equation~(\ref{eq:caract}) can be further simplified
to~(\ref{chnp}) and~(\ref{chni}), by dealing separately with even and odd values of~$N$.


\section*{References}

\begin{thebibliography}{99}

\bibitem{Aharonov} Aharonov Y, Davidovich L and Zagury N 1993
Quantum random walks
{\it Phys. Rev. A} {\bf 48} 1687

\bibitem{Farhi} Farhi E and Gutmann S 1998
Quantum computation and decision trees
{\it Phys. Rev. A} {\bf 58} 915

\bibitem{Childs} Childs A M 2009
Universal computation by quantum walk
{\it Phys. Rev. Lett.} {\bf 102} 180501

\bibitem{Kempe} Kempe J 2003
Quantum random walks -- an introductory overview
{\it Contemp. Phys.} {\bf 44} 307

\bibitem{Ambainis} Ambainis A 2003
Quantum walks and their algorithmic applications
{\it Int. J. Quantum Inf.} {\bf 1} 507

\bibitem{Venegas} Venegas-Andraca S E 2012
Quantum walks: a comprehensive review
{\it Quantum Inf. Process.} {\bf 11} 1015

\bibitem{Ryan} Ryan C A, Laforest M, Boileau J C and Laflamme R 2005
Experimental implementation of a discrete-time quantum random walk on an NMR quantum-information processor
{\it Phys. Rev. A} {\bf 72} 062317

\bibitem{Schmitz} Schmitz H, Matjeschk R, Schneider C, Glueckert J, Enderlein M, Huber T and Schaetz T 2009
Quantum walk of a trapped ion in phase space
{\it Phys. Rev. Lett.} {\bf 103} 090504

\bibitem{Zahringer} Z\"ahringer F, Kirchmair G, Gerritsma R, Solano E, Blatt R and Roos C F 2010
Realization of a quantum walk with one and two trapped ions
{\it Phys. Rev. Lett.} {\bf 104} 100503

\bibitem{Karski} Karski M, F\"orster L, Choi J M, Steffen A, Alt W, Meschede D and Widera A 2009
Quantum walk in position space with single optically trapped atoms
{\it Science} {\bf 325} 174

\bibitem{Schreiber} Schreiber A, Cassemiro K N, Poto\v{c}ek V, G\'abris A, Mosley P J, Andersson E, Jex I and Silber\-horn~C 2010
Photons walking the line: A quantum walk with adjustable coin operations
{\it Phys. Rev. Lett.} {\bf 104} 050502

\bibitem{Knight} Knight P L, Rold\'an E and Sipe J E 2003
Quantum walk on the line as an interference phenomenon
{\it Phys. Rev. A} {\bf 68} 020301(R)

\bibitem{Perets} Perets H B, Lahini Y, Pozzi F, Sorel M, Morandotti R and Silberberg Y 2008
Realization of quantum walks with negligible decoherence in waveguide lattices
{\it Phys. Rev. Lett.} {\bf 100} 170506

\bibitem{Hillery} Hillery M 2010
Quantum walks through a waveguide maze
{\it Science} {\bf 329} 1477

\bibitem{Peruzzo} Peruzzo A, Lobino M, Matthews J C F, Matsuda N, Politi A, Poulios K, Zhou X Q, Lahini~Y, Ismail N, W\"orhoff K, Bromberg Y, Silberberg Y, Thompson M G and OBrien J L 2010
Quantum walk of correlated photons
{\it Science} {\bf 329} 1500

\bibitem{Lahini} Lahini Y, Bromberg Y, Christodoulides D N and Silberberg Y 2010
Quantum correlations in two-particle Anderson localization
{\it Phys. Rev. Lett.} {\bf 105} 163905

\bibitem{Sansoni} Sansoni L, Sciarrino F, Vallone G, Mataloni P, Crespi A, Ramponi R and Osellame R 2012
Two-particle bosonic-fermionic quantum walk via integrated photonics
{\it Phys. Rev. Lett.} {\bf 108} 010502

\bibitem{Omar} Omar Y, Paunkovic N, Sheridan L and Bose S 2006
Quantum walk on a line with two entangled particles
{\it Phys. Rev. A} {\bf 74} 042304

\bibitem{Gamble} Gamble J K, Friesen M, Zhou D, Joynt R and Coppersmith S N 2010
Two-particle quantum walks applied to the graph isomorphism problem
{\it Phys. Rev. A} {\bf 81} 052313

\bibitem{Stefanak} \v{S}tefa\v{n}\'ak M, Barnett S M, Koll\'ar B, Kiss T and Jex I 2011
Directional correlations in quantum walks with two particles
{\it New J. Phys.} {\bf 13} 033029

\bibitem{Chandrashekar} Chandrashekar C M and Busch T 2012
Quantum walk on distinguishable non-interacting many-particles and indistinguishable two-particle
{\it Quantum Inf. Process.} {\bf 11} 1287

\bibitem{Andrei} Qin X, Ke Y, Guan X W, Li Z, Andrei N and Lee C 2014
Statistics-dependent quantum co-walking of two particles in one-dimensional lattices with nearest-neighbor interaction
{\it Phys. Rev. A} {\bf 90} 062301

\bibitem{ToroJML} de Toro Arias S and Luck J M 1998
Anomalous dynamical scaling and bifractality in the one-dimensional Anderson model
{\it J. Phys. A: Math. Gen.} {\bf 31} 7699

\bibitem{KLM} Krapivsky P L, Luck J M and Mallick K 2015
Interacting quantum walkers: Two-body bosonic and fermionic bound states
{\it J. Phys. A: Math. Theor.} {\bf 48} 475301

\bibitem{Tibor} Antal T, Krapivsky P L and Mallick K 2007
Molecular spiders in one dimension
{\it J. Stat. Mech.} {\bf 2007} P08027

\bibitem{Baraviera} Baraviera A, Franco T and Neumann A 2015
Hydrodynamic limit of quantum random walks
in {\it From Particle Systems to Partial Differential Equations II} Springer Proceedings in Mathematics \& Statistics {\bf 129} [arXiv:1309.1146]

\bibitem{Grimmett} Grimmett G, Janson S and Scudo P F 2004
Weak limits of quantum random walks
{\it Phys. Rev. E} {\bf 69} 026119

\bibitem{Gottlieb} Gottlieb A D 2005
Convergence of continuous-time quantum walks on the line
{\it Phys. Rev. E} {\bf 72} 047102

\bibitem{Konno} Konno N 2005
Limit theorem for continuous-time quantum walk on the line
{\it Phys. Rev. E} {\bf 72} 026113

\bibitem{Strauch} Strauch F W 2006
Connecting the discrete- and the continuous-time quantum walk
{\it Phys. Rev. A} {\bf 74} 030301(R)

\bibitem{r1} M\"ulken O, Pernice V and Blumen A 2008
Universal behavior of quantum walks with long-range steps
{\it Phys. Rev. E} {\bf 77} 021117

\bibitem{r2} M\"ulken O and Blumen A 2011
Continuous-time quantum walks: Models for coherent transport on complex networks
{\it Phys. Rep.} {\bf 502} 37

\bibitem{x1} Xu X P 2008
Continuous-time quantum walks on one-dimensional regular networks
{\it Phys. Rev. E} {\bf 77} 061127

\bibitem{x2} Xu X P 2009
Coherent exciton transport and trapping on long-range interacting cycles
{\it Phys. Rev. E} {\bf 79} 011117

\bibitem{g1} Geim A K and Novoselov K S 2007
The rise of graphene
{\it Nature Materials} {\bf 6} 183

\bibitem{g2} Castro-Neto A H, Guinea F, Peres N M R, Novoselov K S and Geim A K 2009
The electronic properties of graphene
{\it Rev. Mod. Phys.} {\bf 81} 109

\bibitem{krbbook} Krapivsky P L, Redner S and Ben-Naim E 2010
{\it A Kinetic View of Statistical Physics}
(Cambridge: Cambridge University Press)

\bibitem{Nepomechie} Nepomechie R I 2001
Bethe Ansatz solution of the open XX spin chain with non-diagonal boundary terms
{\it J. Phys. A: Math. Gen.} {\bf 34} 9993

\bibitem{Samaj} \v{S}amaj L and Bajnok J 2013
{\it Introduction to the Statistical Physics of Integrable Many-body Systems}
(Cambridge: Cambridge University Press)

\bibitem{Lieb} Lieb E, Schultz T and Mattis D 1961
Two soluble models of an antiferromagnetic chain
{\it Ann. Phys.} {\bf 16} 407

\bibitem{Hinrichsen} Hinrichsen H, Krebs K and Peschel I 1996
Solution of a one-dimensional diffusion-reaction model with spatial asymmetry
{\it Z. Phys. B} {\bf 100} 105

\bibitem{Birgit} Bilstein U and Wehefritz B 1999
The XX-model with boundaries: Part I. Diagonalisation of the finite chain
{\it J. Phys. A: Math. Gen.} {\bf 32} 191

\bibitem{GR} Gradshteyn I S and Ryzhik I M 1965
{\it Table of Integrals, Series, and Products}
(New York: Academic)

\bibitem{exp} Pei R, Taylor S K, Stefanovic D, Rudchenko S, Mitchell T E and Stojanovic M N 2006
Behavior of polycatalytic assemblies in a substrate-displaying matrix
{\it J. Am. Chem. Soc.} {\bf 128} 12693

\bibitem{iks} Inui N, Konno N and Segawa E 2005
One-dimensional three-state quantum walk
{\it Phys. Rev. E} {\bf 72} 056112

\bibitem{sbj} \v{S}tefa\v{n}\'ak M, Bezdekova I and Jex I 2014
Limit distributions of three-state quantum walks: The role of coin eigenstates
{\it Phys. Rev. A} {\bf 90} 012342

\bibitem{str1} Brouwer P W, Mudry C, Simons B D and Altland A 1998
Delocalization in coupled one-dimensional chains
{\it Phys. Rev. Lett.} {\bf 81} 862

\bibitem{str2} Mudry C, Brouwer P W and Furusaki A 1999
Random magnetic flux problem in a quantum wire
{\it Phys. Rev. B} {\bf 59} 13221

\bibitem{str3} Mudry C, Brouwer P W and Furusaki A 1999
Crossover from the chiral to the standard universality classes in the conductance of a quantum wire with random hopping only
{\it Phys. Rev. B} {\bf 62} 8249

\bibitem{str4} Brouwer P W, Mudry C and Furusaki A 2000
Nonuniversality in quantum wires with off-diagonal disorder: a geometric point of view
{\it Nucl. Phys. B} {\bf 565} 653

\end{thebibliography}
\end{document}